\documentclass[reprint,aps,prc,amsmath,amssymb,citeautoscript,showkeys,floatfix,superscriptaddress,nofootinbib,nobalancelastpage]{revtex4-2}

\usepackage[T1]{fontenc}
\usepackage{braket}
\usepackage{mathtools}
\usepackage{diffcoeff}[=v4]
\usepackage{bm}
\usepackage{hyperref}
\usepackage{isotope}
\usepackage{natbib}
\usepackage{graphicx}
\usepackage{siunitx}
\usepackage{xspace}
\usepackage{dcolumn}
\usepackage{booktabs}

\renewcommand{\vec}[1]{\bm{\mathbf{#1}}}
\renewcommand{\equiv}{\stackrel{\mathrm{def}}{=}}

\newcommand{\grad}{\vec{\nabla}}
\newcommand{\uvec}[1]{\vec{\hat{#1}}}
\newcommand{\NnLO}[1]{N\textsuperscript{#1}LO}
\newcommand{\LO}{LO}
\newcommand{\NLO}{NLO}
\newcommand{\aN}[1]{#1{N}}
\newcommand{\order}{{}}
\newcommand{\intd}[1]{\int\displaylimits#1}

\newcommand{\Nmax}{{N_\text{max}}}
\newcommand{\hw}{{\hbar\omega}}
\makeatletter
\@ifclasswith{revtex4-2}{onecolumn}{\newcommand{\ifproofpre}[2]{#2}}{\newcommand{\ifproofpre}[2]{#1}}
\newenvironment{maybemultline}{\ifproofpre{\multline}{\equation}}{\ifproofpre{\endmultline}{\endequation}}
\makeatother

\DeclareRobustCommand{\tme}[3]{\langle {#1} \vert {#2} \vert {#3} \rangle}
\DeclareRobustCommand{\trme}[3]{\langle #1 \Vert #2 \Vert #3 \rangle}
\newcommand{\tcgseparator}{}
\DeclareRobustCommand{\tcg}[6]{(#1#2\tcgseparator#3#4\vert#5#6)}
\DeclareRobustCommand{\toverlap}[2]{\langle #1 \vert #2 \rangle}

\DeclareSIUnit[number-unit-product={}]{\nuclearmagneton}{\ensuremath{\,\mu_N}}

\begin{document}

\title{Magnetic moments of \(A = 3\) nuclei with chiral effective field theory operators}

\author{Soham Pal}
\altaffiliation[Present address: ]{Research and Discovery Technologies, University of Arizona, Tucson, AZ 85721-0073}
\affiliation{Department of Physics and Astronomy, Iowa State University, Ames, Iowa 50011-3160, USA}

\author{Shiplu Sarker}
\affiliation{Department of Physics and Astronomy, Iowa State University, Ames, Iowa 50011-3160, USA}

\author{Patrick J. Fasano}
\altaffiliation[Present address: ]{Physics Division, Argonne National Laboratory, Argonne, Illinois 60439-4801, USA}
\affiliation{Department of Physics and Astronomy, University of Notre Dame, Notre Dame, Indiana 46556-5670, USA}

\author{Pieter Maris}
\affiliation{Department of Physics and Astronomy, Iowa State University, Ames, Iowa 50011-3160, USA}

\author{James P. Vary}
\affiliation{Department of Physics and Astronomy, Iowa State University, Ames, Iowa 50011-3160, USA}

\author{Mark A. Caprio}
\affiliation{Department of Physics and Astronomy, University of Notre Dame, Notre Dame, Indiana 46556-5670, USA}

\author{Robert A. M. Basili}
\affiliation{Department of Electrical and Computer Engineering, Iowa State University, Ames, Iowa 50011-3160, USA}

\date{\today}

\begin{abstract}
Chiral effective field theory ($\chi$EFT)
provides a framework for obtaining internucleon interactions in a systematically
improvable fashion from first principles, while also providing for the
derivation of consistent electroweak current operators.  In this work, we apply
consistently derived interactions and currents towards calculating the magnetic
dipole moments of the $A=3$ systems $\isotope[3]{H}$ and $\isotope[3]{He}$.  We
focus here on LENPIC interactions obtained using semilocal coordinate-space
(SCS) regularization.  Starting from the momentum-space representation of the
LENPIC $\chi$EFT vector current, we derive the SCS-regularized magnetic dipole
operator up through \NnLO{2}.  We then carry out no-core shell model calculations for
$\isotope[3]{H}$ and $\isotope[3]{He}$ systems, using the SCS LENPIC interaction
at \NnLO{2} in $\chi$EFT, and evaluate the magnetic dipole
moments obtained using the consistently derived one-nucleon and two-nucleon
electromagnetic currents.  As anticipated by prior results with $\chi$EFT
currents, the current corrections through \NnLO{2} provide improved, but not yet
complete, agreement with experiment for the $\isotope[3]{H}$ and
$\isotope[3]{He}$ magnetic dipole moments.
\end{abstract}

\maketitle

\section{Introduction}
\label{Introduction}

Chiral effective field theory ($\chi$EFT) is a systematically improvable
approach to obtain internucleon interactions and corresponding electroweak
current operators from first
principles~\cite{weinberg-1990-nuclear-forces,weinberg-1991-effective,ordonez-1992-chiral,ordonez-1994-nucleon-nucleon}.
Multiple implementations of
$\chi$EFT have emerged that differ in the choice of the subnuclear degrees of freedom, power
counting scheme, and choice of regulators.  This has led to several internucleon
interactions that accurately describe nucleon-nucleon scattering data and
the deuteron bound state.  Under each such implementation of $\chi$EFT, corresponding electroweak
current operators may be derived, subject to various challenges in obtaining
consistency~\cite{krebs2020:nuclear-currents}.

The present work is focused on the next stage in the process, where we apply
consistently derived interactions and currents towards calculating nuclear
physics observables.  Historically, corrections to the naive electroweak
operators were obtained phenomenologically from meson-exchange
theory~\cite{friar1977:pion-exchange,friar1980:deuteron-charge}.
The program of developing electroweak currents from
$\chi$EFT was initiated in the context of hybrid approaches which combined
phenomenological internucleon interactions with incomplete $\chi$EFT
currents~\cite{park-1993-chiral,
  park-1996-chiral-lagrangian,park-2001-effective-field,
  park-2003-parameter-free, song-2007-effective-field-theory}.

A new generation of $\chi$EFT interactions, and their corresponding
currents, have been derived from $\chi$EFT by constructing effective operators which
act only on nucleonic degrees of freedom, either by the method of unitary
transformations (UT)~\cite{eden1996:3n-hard-core,epelbaum-1998-nuclear-1}
or by means of time-ordered perturbation theory
(TOPT)~\cite{weinberg-1990-nuclear-forces,weinberg-1991-effective,ordonez-1992-chiral}.
For the Norfolk \(\chi\)EFT potentials~\cite{piarulli-2014-minimally, piarulli-2016-local},
which are local and include $\Delta$ intermediate states, the energy-dependence resulting from the application
of TOPT is removed through an inverse $T$-matrix
approach~\cite{pastore2011:em-charge-n4lo,piarulli2013:electromagnetic-structure}.
For the potentials of the Low Energy Nuclear Physics International Collaboration
(LENPIC)~\cite{epelbaum-2015-precision-nucleon,epelbaum-2015-improved,Reinert:2020mcu},
which are nonlocal and include only pion intermediate states, operators acting purely on nucleonic
degrees of freedom are constructed
using the UT method~\cite{epelbaum-1998-nuclear-1,epelbaum-1999-nuclear-2}.

In this work, we apply consistently derived interactions and currents towards
calculating the magnetic dipole moments of the $A=3$ systems $\isotope[3]{H}$
and $\isotope[3]{He}$.  The magnetic dipole moments of these systems have
previously been calculated using \(\chi\)EFT currents, both in hybrid approaches
with phenomenological potentials~\cite{song-2007-effective-field-theory,
  song-2009-up, pastore-2013-quantum-monte}, and in a fully \(\chi\)EFT approach
using the Norfolk potentials and currents~\cite{schiavilla-2019-local}.  We
focus here on LENPIC interactions obtained using semilocal coordinate-space
(SCS) regularization, developed to preserve the approximately-local nature of
the long-range potentials, and associated currents. We calculate the magnetic
dipole moments using wave functions obtained by no-core shell model
(NCSM)~\cite{barrett-2013-ab} calculations. We use the SCS regularized two-nucleon ($2N$) and
two-nucleon plus three-nucleon ($2N + 3N$) LENPIC potentials up to next-to-next-to-leading order (\NnLO{2}), with the consistently derived single-nucleon ($1N$)
and $2N$ electromagnetic currents. Consequences of applying a similarity
renormalization group (SRG)
transformation~\cite{glazek-1993-renormalization-hamiltonians,
  glazek-1994-perturbative-renormalization, wegner-1994-flow} to the potential
are also considered.  Initial results were reported in Ref.~\cite{pal2022:diss}.

We first derive the SCS-regularized magnetic dipole operator starting from the
momentum-space representation of the LENPIC $\chi$EFT vector current
(Sec.~\ref{sec:derivation}). We then detail the calculational scheme used for
our NCSM calculations of magnetic dipole moments for the $A=3$ systems
(Sec.~\ref{sec:calculations}), and present our results for the magnetic dipole
moments obtained with the LENPIC SCS-regulated interaction and $\chi$EFT
magnetic dipole operator through \NnLO{2} (Sec.~\ref{sec:results}).  We discuss
the effects of including the $3N$ interaction and of the choice of SCS regulator
parameter and SRG evolution, and compare to prior results.  In order to achieve
a compact presentation, we include most of the formal developments, regarding
the derivation of the SCS-regularized operators, in Appendices.

\section{Magnetic dipole operator from \(\chi\)EFT}
\label{sec:derivation}

The magnetic dipole moment, characterizing the interaction between a charged current
and the electromagnetic field, is defined
classically as~\cite{jackson1999:classical-electrodynamics}
\begin{equation}
    \label{eq:dipole-moment-coordinate}
    \vec{\mu} = \frac{1}{2} \int_{\mathbb{R}^3} \dl^3 x \, \vec{x} \times \bar{\vec{j}}(\vec{x}),
\end{equation}
where $\bar{\vec{j}}(\vec{x})$ is the charged-current density at position $\vec{x}$.
We can also express the magnetic dipole moment in terms of the current density in momentum space
via the Fourier transform,
\begin{equation}
    \bar{\vec{j}}(\vec{x}) = \int_{\mathbb{R}^3} \frac{\dl^3 k}{(2\pi)^{3/2}} \, e^{i \vec{k}\cdot\vec{x}} \vec{j}(\vec{k}),
\end{equation}
obtaining
\begin{equation}
    \label{eq:dipole-moment-momentum}
    \vec{\mu} = \frac{1}{2i} \vec{\nabla}_{\vec{k}} \times \vec{j}(\vec{k}) \Big\vert_{\vec{k}=0}.
\end{equation}

If we
take the momentum-space matrix element of the non-relativistic current
operator~\cite{krebs2019:n4lo-em} for a charged particle with spin, we obtain
\begin{equation}
    \vec{j}(\vec{p}', \vec{p}; \vec{k}) =2\mu_N \left(
        g_{l} \frac{\vec{p}'+\vec{p}}{2}
        - i g_{s} \vec{k}\times\vec{s}
    \right),
\end{equation}
with the connection to the quantum mechanical matrix element given by
\begin{equation}
  \label{eq:j-me-def}
    \tme{\vec{p}'}{\vec{j}(\vec{k})}{\vec{p}} \equiv \vec{j}(\vec{p}', \vec{p}; \vec{k}) \delta^3(\vec{p}' - \vec{p} - \vec{k})
\end{equation}
following the notation of Ref.~\cite{krebs2017:n4lo-axial},
where $\mu_N = e/2m_N$ is the nuclear magneton, $\vec{p}$ ($\vec{p}'$) is the
initial (final) momenta, $\vec{s}$ is the spin operator, \(g_{l}\) and
\(g_{s}\) are the orbital and spin \(g\) factors, and the momentum eigenstates are normalized as $\toverlap{\vec{p}'}{\vec{p}} = \delta^{(3)}(\vec{p}'-\vec{p})$.
Combining this with \eqref{eq:dipole-moment-momentum} and
writing the orbital angular momentum
$\vec{l} = - \vec{p} \times i \vec{\nabla}_{\vec{p}}$ in momentum space, we get the
conventional~\cite{bohr-1998-nuclear-structure,suhonen-2007-nucleons-to-nucleus}
(impulse approximation) expression for the magnetic dipole moment operator\footnote{The magnetic dipole operator ${\vec{\mu}}$ considered here, normalized
  appropriately for calculation of the magnetic dipole moment, is related to the
  magnetic dipole operator ${\vec{M}}_1$ found in the theory of electromagnetic
  transitions (see Appendix~\ref{sec:magn-mult-oper}) by a conventional factor,
  as ${\vec{\mu}} = (4\pi/3)^{1/2}{\vec{M}}_1$.
}\begin{equation}
  \label{eq:mu-ia-def}
  \vec{\mu}^\mathrm{IA} = \mu_N  \left(g_{l} \vec{l} + g_{s} \vec{s}\right),
\end{equation}
where \(\vec{l}\) and \(\vec{s}\) are the orbital and spin angular momentum
operators, respectively. The operator
$\vec{\mu}^\mathrm{IA}$ is a one-body operator which corresponds to the treatment of
nucleons as point particles with charges and intrinsic magnetic moments.

For the present work, we have used the LENPIC SCS-regulated potentials described in
Refs.~\cite{epelbaum-2015-improved,epelbaum-2015-precision-nucleon,epelbaum-2019-few-many},
and previously used for low-energy nuclear structure calculations
in Refs.~\cite{binder-2016-few, binder-2018-few,epelbaum-2019-few-many}.
The $2N$ potentials have been derived
up to \NnLO{4} in the chiral order and fitted to nucleon-nucleon scattering data
and the deuteron bound state, while the $3N$ interactions have
been derived up to \NnLO{2} and fitted to nucleon-deuteron ($Nd$) scattering.

Because iteration of the $2N$ interaction with the Lippmann-Schwinger equation
generates ultraviolet (UV) divergences~\cite{epelbaum2009:regularization,machleidt2011:chiral-eft-nuclear},
one must regulate the high momentum
(or, equivalently, short distance) behavior of the interaction.
This is usually  done by introducing a momentum-space UV cutoff \(\Lambda\). Choosing a large value for
\(\Lambda\), such as the mass of the \(\rho\) meson, results in spurious deeply
bound states, while choosing a small cutoff leads to more-pronounced
finite-cutoff artifacts (for more details, see
Ref.~\cite{epelbaum-2015-improved}).

To attempt to mitigate finite-cutoff artifacts,
in the LENPIC SCS framework, a hybrid regularization
scheme has been adopted.
The terms in these potentials arising from pion exchange (without contact interactions)
have been regularized in coordinate space by multiplying with
the coordinate space function
\begin{equation}
  \label{eq:local-regulator-function}
  f(r) = \left[1 - \exp\left(-\frac{r^2}{R^2}\right)\right]^6,
\end{equation}
where $r$ is the relative separation between the two nucleons, and $R$ characterizes
the cutoff separation.
Meanwhile, the contact terms have been regularized in momentum space by multiplying by
the nonlocal Gaussian regulator
\begin{equation}
    \label{eq:nonlocal-regulator-function}
    g(p,p') = \exp \left(-\frac{p^2 + p'^2}{\Lambda^2}\right),
\end{equation}
where $p$ and $p'$ are the magnitudes of the incoming and outgoing relative nucleon momenta, respectively,
with the cutoff \(\Lambda = 2R^{-1}\). Here we
have two sets of interactions, one set with
\(R = \qty{0.9}{\femto\meter}\) and the other with
\(R = \qty{1.0}{\femto\meter}\).

For consistency with the regularization scheme for the interaction, we must
also regularize the operators that arise from the $\chi$EFT expansion of the
magnetic dipole moment operator. The current-density operator is typically derived
and expressed in momentum space~\cite{koelling-2011-two,krebs2019:n4lo-em,krebs2020:nuclear-currents}.
However, in order to apply the regulator function in \eqref{eq:local-regulator-function},
we must transform the long-range parts  of the magnetic dipole moment operator to coordinate space.
For the coordinate-space matrix element of the $a$-body operator
${\vec{\mu}}^{\aN{a}}_{\order}$
we have~\cite{bohr-1998-nuclear-structure, carlson-1998-structure}
\begin{maybemultline}
\label{eq:mu-general-def}
  \vec{\mu}^{\aN{a}}_{\order}(\vec{r}'_1, \ldots, \vec{r}'_a, \vec{r}_1, \ldots, \vec{r}_a)
  \ifproofpre{\\}{}
    = \frac{1}{2}
    \int \dl^3 x\, \vec{x} \times \bar{\vec{j}}^{\aN{a}}_{\order}(\vec{r}'_1, \ldots, \vec{r}'_a, \vec{r}_1, \ldots, \vec{r}_a; \vec{x}),
\end{maybemultline}
where \(\vec{r}_i\, (\vec{r}_i')\) is the initial (final) position of the \(i\)th
nucleon.
We define the coordinate-space matrix element of $\bar{\vec{j}}^{\aN{a}}$ via
\begin{equation}
   \tme{\vec{r}'_1 \cdots \vec{r}'_a}{\bar{\vec{j}}(\vec{x})}{\vec{r}_1 \cdots \vec{r}_a} = \bar{\vec{j}}(\vec{r}'_1, \ldots, \vec{r}'_a, \vec{r}_1, \ldots, \vec{r}_a; \vec{x}).
\end{equation}
In order to use the (momentum-space) current matrix elements
derived in Refs.~\cite{koelling-2011-two} and \cite{krebs2019:n4lo-em} with the coordinate-space regulators, we
perform the change of basis via the multidimensional Fourier transform
\begin{multline}
  \label{eq:current-fourier-transformation}
  \bar{\vec{j}}^{\aN{a}}_{\order}(\vec{r}'_1, \ldots, \vec{r}'_a, \vec{r}_1, \ldots, \vec{r}_a; \vec{x})
\\
  \quad
    = \intd_{\{\vec{q}\}_1^{a}}
    \intd_{\{\vec{Q}\}_1^{a}}
    \intd_{\vec{k}}
\prod_{i=1}^a
    e^{i\vec{q}_i \cdot (\vec{r}_i' + \vec{r}_i) / 2}\,
e^{i\vec{Q}_i \cdot \Delta\vec{r}_i}\,
e^{i\vec{k} \cdot \vec{x}}\,
\ifproofpre{\\ \times}{}
    \vec{j}^{\aN{a}}_{\order}(\vec{q}_1, \ldots, \vec{q}_a, \vec{Q}_1, \ldots, \vec{Q}_a; \vec{k})
\\ \times
    (2\pi)^{-3a+3}\delta^{(3)}(\vec{q}_1 + \cdots + \vec{q}_a - \vec{k}),
\end{multline}
where
\(\vec{q}_i = \vec{p}_i' - \vec{p}_i\),
\(\vec{Q}_i = (\vec{p}_i' + \vec{p}) / 2\) are linear combinations of the
incoming (\(\vec{p}_i\)) and outgoing (\(\vec{p}_i'\)) momenta of the \(i\)th
nucleon, and \(\vec{k}\) is the momentum of the external electromagnetic
field.
Following the convention of Refs.~\cite{krebs2017:n4lo-axial,krebs2019:n4lo-em}, we
define the function
$\vec{j}^{\aN{a}}(\vec{q}_1,\ldots,\vec{q}_a,\vec{Q}_1,\ldots,\vec{Q}_a;\vec{k})$
in terms of the momentum space matrix element
\begin{maybemultline}
    \label{eq:moment-space-normalization-a-body}
    \tme{\vec{p}'_1 \cdots \vec{p}'_a}{{\vec{j}}(\vec{k})}{\vec{p}_1 \cdots \vec{p}_a}
        \ifproofpre{\\}{}
    = (2\pi)^{-3a+3} \delta^{(3)}(\vec{q}_1+\cdots+\vec{q}_a - \vec{k})
        \ifproofpre{\\ \times}{}
    \vec{j}(\vec{q}_1, \ldots, \vec{q}_a, \vec{Q}_1, \ldots, \vec{Q}_a; \vec{k}),
\end{maybemultline}
where we adopt the non-relativistic normalization of states
$\toverlap{\vec{p}'}{\vec{p}} = \delta^{(3)}(\vec{p}'-\vec{p})$.
We also use the notations
\(\intd_{\{\vec{q}\}_{1}^{a}} = \intd_{\vec{q}_{1}} \cdots \intd_{\vec{q}_{a}}\),
with \(\intd_{\vec{q}} = \int \frac{\dl^3 q}{(2\pi)^{3/2}}\),
and
\(\Delta\vec{r}_i = \vec{r}_i' - \vec{r}_i\).
From equations
\eqref{eq:mu-general-def} and \eqref{eq:current-fourier-transformation} we obtain
\begin{multline}
  \label{eq:mu-general-def-momentum}
  \vec{\mu}^{\aN{a}}_\order(\ldots)
  \ifproofpre{\\}{}
    = \frac{(2\pi)^{3}}{2i}
    \Biggl[
    \grad_{\vec{k}} \times
    \intd_{\{\vec{q}\}_1^{a}}
    \intd_{\{\vec{Q}\}_1^{a}}
    \prod_{i=1}^a
    e^{i\vec{q}_i \cdot (\vec{r}_i' + \vec{r}_i) / 2}\,
    e^{i\vec{Q}_i \cdot \Delta\vec{r}_i}\,
    \vec{j}^{\aN{a}}_\order(\ldots)
  \ifproofpre{\\ \times}{}
    \delta^{(3)}(\vec{q}_1 + \cdots + \vec{q}_a - \vec{k})
    \Biggr]_{\vec{k}=0},
\end{multline}
where for brevity we have omitted the arguments of
\(\vec{\mu}^{\aN{a}}_{\order}\) and \(\vec{j}^{\aN{a}}_{\order}\).

We use these relations to derive the magnetic dipole operators from the
corresponding momentum space electromagnetic currents. While we will now focus
on specific $1N$ and $2N$ currents, we emphasize that this relation is
true for any electromagnetic current derived from \(\chi\)EFT. For a general
derivation of electric and magnetic multipole operators see appendix
\ref{sec:em-mult-operators}.

\subsection{Single-nucleon magnetic dipole operators \label{sec:single-nucl-magn}}
For 1N currents, after integrating over \(\vec{q}_1\) and expanding
the curl, equation~\eqref{eq:mu-general-def-momentum} reduces to
\begin{multline}
  \label{eq:mu-1N-generic}
  \vec{\mu}^{\aN{1}}_{\order} = \frac{1}{2i}
    \Biggl[
    \frac{i}{2}(\vec{r}_1' + \vec{r}_1) \times
    \intd_{\vec{Q}_1} e^{i\vec{Q}_1 \cdot \Delta\vec{r}_1}\,
    \vec{j}^{\aN{1}}_{\order}(\vec{q}_1 = \vec{k}, \vec{Q}_1)
    \ifproofpre{\\}{}
    +
    \intd_{\vec{Q}_1} e^{i\vec{Q}_1 \cdot \Delta\vec{r}_1}\,
    \grad_{\vec{k}} \times \vec{j}^{\aN{1}}_{\order}(\vec{q}_1 = \vec{k}, \vec{Q}_1)
    \Biggr]_{\vec{k}=0}.
\end{multline}
If the current is independent of \(\vec{Q}_1\) then integrating over
\(\vec{Q}_1\) gives an additional delta function,
\(\delta^{(3)}(\Delta\vec{r}_1)\).
Up to \NnLO{2} in the power counting scheme established in
\cite{krebs2019:n4lo-em}, there are two 1N currents -- one
at next-to-leading order (NLO), and the other at \NnLO{2}. The 1N current at NLO is
\begin{align}
  \label{eq:j-1N-NLO}
  \vec{j}^{\aN{1}}_{\mathrm{NLO}}
= \frac{e}{4m_N}
  \Bigl[
  -i \vec{q}_1 \times \vec{\sigma}_1 (g_s + g_v \tau_{1}^3)
  + 2 \vec{Q}_1 (1 + \tau_{1}^3)
  \Bigr],
\end{align}
where \(m_N\) is the average nucleon mass, \(g_s=\tfrac{1}{2}(g_{s,p}+g_{s,n})\) and \(g_v=\tfrac{1}{2}(g_{s,p}-g_{s,n})\) are the
isoscalar and isovector $g$ factors of the nucleon, respectively, while \(\vec{\sigma}\) and \(\vec{\tau}\) are the Pauli
matrices in spin and isospin spaces, respectively. Substituting this current into
equation~\eqref{eq:mu-1N-generic} we get
\begin{maybemultline}
\label{eq:mu-1N-NLO}
  \vec{\mu}^{\aN{1}}_{\mathrm{NLO}} = \frac{\mu_N}{2}
  \Bigl[
  (g_s + g_v \tau_{1}^3) \vec{\sigma}_1
  + (1 + \tau_{1}^3) \vec{l}_1
  \Bigr]
  \delta^{(3)}(\Delta\vec{r}_1),
\end{maybemultline}
where
\(\Delta\vec{r}_{1} = \vec{r}_{1}' - \vec{r}_{1}\). This expression is
equivalent to \(\vec{\mu}^\mathrm{IA}\) in equation~\eqref{eq:mu-ia-def}.
The $1N$ current at \NnLO{2}, which arises due to the chiral expansion of the $1N$ form factors, is given by
\begin{align}
\label{eq:j-1N-N2LO}
  \vec{j}^{\aN{1}}_{\text{\NnLO{2}}}
  = - \frac{i eg_{A}^{2}}{32 \pi F_{\pi}^{2} }\tau_{1}^{3}
  \Bigl[
  m_{\pi} - (4 m_{\pi}^{2} + q_{1}^{2}) A(|\vec{q}_{1}|)
  \Bigr]
  (\vec{q}_{1} \times \vec{\sigma}_{1}),
\end{align}
where \(A({q}) = \frac{1}{2q} \tan^{-1}( \frac{q}{m_{\pi}})\), \(g_A\) is the
axial coupling constant, \(m_\pi\) is the average pion mass, and \(F_\pi\) is
the pion decay constant. The chiral expansion of the $1N$ form factors
converges slowly, and so in this work we use physical values of the $1N$ form factors, which at \(\vec{k} = 0\) are just the isoscalar and isovector magnetic
moments in equation (\ref{eq:j-1N-NLO}). This current does not contribute
to the magnetic dipole moment operator, as can be shown by substituting $\vec{j}^{\aN{1}}_{\mathrm{N2LO}}$
into equation~\eqref{eq:mu-general-def-momentum} to obtain zero.

\subsection{Two-nucleon magnetic moment operators \label{sec:two-nucl-magn}}
We define the initial relative and center of mass coordinates for a $2N$
system:
\begin{align}
  \label{eq:rel-cm-coordinates}
  \vec{r}_{12} = \vec{r}_1 - \vec{r}_2, \quad
  \vec{R}_{12} = (\vec{r}_1 + \vec{r}_2) / 2.
\end{align}
The final relative and center of mass coordinates are similarly defined with
\(\vec{r}'_1\) and \(\vec{r}'_2\). For $2N$ currents, after expressing
the nucleon coordinates in terms of these new coordinates, integrating over
\(\vec{q}_2\), and expanding the curl, equation
\eqref{eq:mu-general-def-momentum} reduces to
\begin{widetext}
\begin{multline}
  \label{eq:mu-2N-generic}
  \vec{\mu}^{\aN{2}}_\order
  = \frac{1}{2i} \Biggl[\frac{i}{2} (\vec{R}'_{12} + \vec{R}_{12})
    \\\times 
      \intd_{\vec{q}} \intd_{\{\vec{Q}\}_1^{2}}
      e^{i\vec{q} \cdot (\vec{r}'_{12} + \vec{r}_{12}) / 2}
    e^{i(\vec{Q}_1 + \vec{Q}_2) \cdot \Delta\vec{R}_{12}}
e^{i(\vec{Q}_1 - \vec{Q}_2) \cdot \Delta\vec{r}_{12} / 4}
    \vec{j}^{\aN{2}}_\order (
    \tfrac{1}{2}\vec{k} + \vec{q},
    \tfrac{1}{2}\vec{k} -\vec{q},
    \vec{Q}_1, \vec{Q}_2)
    \\
  \qquad
    + \intd_{\vec{q}} \intd_{\{\vec{Q}\}_1^{2}}
    e^{i\vec{q} \cdot (\vec{r}'_{12} + \vec{r}_{12}) / 2}
    e^{i(\vec{Q}_1 + \vec{Q}_2) \cdot \Delta\vec{R}_{12}}
    e^{i(\vec{Q}_1 - \vec{Q}_2) \cdot \Delta\vec{r}_{12} / 4}
    \grad_{\vec{k}} \times
    \vec{j}^{\aN{2}}_\order
    (
    \tfrac{1}{2}\vec{k} + \vec{q},
    \tfrac{1}{2}\vec{k} -\vec{q}, \vec{Q}_1, \vec{Q}_2
    )
    \Biggr]_{\vec{k} = 0},
\end{multline}
\end{widetext}
where we have replaced \(\vec{q}_{1} \rightarrow \tfrac{1}{2}\vec{k}+\vec{q}\) and
\(\vec{q}_{2} \rightarrow \tfrac{1}{2}\vec{k}-\vec{q}\) in the arguments to the current $\vec{j}^{\aN{2}}_\order$,
while \(\Delta\vec{r}_{12} = \vec{r}'_{12} - \vec{r}_{12}\) and
\(\Delta\vec{R}_{12} = \vec{R}'_{12} - \vec{R}_{12}\).
If the current is independent of \(\vec{Q}_1\) and \(\vec{Q}_2\) then we get
additional delta functions, \(\delta^{(3)}(\Delta\vec{R}_{12})
\delta^{(3)}(\Delta\vec{r}_{12})\).

Up to \NnLO{2} there is one $2N$ current, arising from the seagull and
pion-in-flight diagrams at \NLO{} \cite{koelling-2011-two}:
\begin{maybemultline}
  \label{eq:j-2N-NLO}
  \vec{j}^{\aN{2}}_{\mathrm{NLO}}
  = \frac{ieg_A^2}{4F_\pi^2}[\vec{\tau}_1 \times \vec{\tau}_2]^3
  \frac{\vec{\sigma}_2 \cdot \vec{q}_2}{q_2^2 + m_\pi^2}
    \Biggl( \vec{q}_1 \frac{\vec{\sigma}_1 \cdot \vec{q}_1}{q_1^2 + m_\pi^2} -
    \vec{\sigma}_1 \Biggr)
  \ifproofpre{\\}{}
    + 1 \leftrightharpoons 2
\end{maybemultline}
Using (\ref{eq:mu-2N-generic}) we get the associated magnetic dipole operator
\begin{maybemultline}
  \label{eq:mu-2N-NLO-Rr}
  \vec{\mu}^{\aN{2}}_{\mathrm{NLO}}
    =
    g_{\pi}
    [\vec{\tau}_1 \times \vec{\tau}_2]^3
    \Bigl[\vec{\mu}^{\aN{2}}_{\mathrm{NLO, cm\text{-}dep}}
    (\vec{R}_{12}, \vec{r}_{12})
  \ifproofpre{\\}{}
    + \vec{\mu}^{\aN{2}}_{\mathrm{NLO, cm\text{-}indep}}
    (\vec{r}_{12})\Bigr]
    \delta^{(3)}(\Delta\vec{R}_{12})
    \delta^{(3)}(\Delta\vec{r}_{12}),
\end{maybemultline}
where the center of mass dependent part is
\begin{multline}
  \label{eq:mu-2N-NLO-Rr-cmdep}
  \vec{\mu}^{\aN{2}}_{\mathrm{NLO, cm\text{-}dep}}
    (\vec{R}_{12}, \vec{r}_{12})
  \ifproofpre{\\}{}
    = \uvec{R}_{12} \times \uvec{r}_{12}
    (m_\pi R_{12})
    Y_0(z)
  \ifproofpre{\\ \times}{}
    [
    Y_2(z)
    \vec{\sigma}_1 \cdot \uvec{r}_{12} \
    \vec{\sigma}_2 \cdot \uvec{r}_{12}
    - Y_1(z)
    \vec{\sigma}_1 \cdot \vec{\sigma}_2
    ]
    ,
\end{multline}
and the center of mass independent part is
\begin{maybemultline}
  \label{eq:mu-2N-NLO-Rr-cmindep}
  \vec{\mu}^{\aN{2}}_{\mathrm{NLO, cm\text{-}indep}}
    (\vec{r}_{12})
  \ifproofpre{\\}{}
    =
    [
    (1 + z)
    (\vec{\sigma}_1 \times \vec{\sigma}_2)
    \cdot \uvec{r}_{12}
    \uvec{r}_{12}
    -z (\vec{\sigma}_1 \times \vec{\sigma}_2)
    ]
    Y_0(z),
\end{maybemultline}
where a hat on a symbol denotes a unit vector,
\(g_{\pi} = -\frac{2 m_N}{e} \frac{e g_A^2 m_\pi}{32 \pi F_\pi^2}\),
\(Y_2(z) = z + \frac{3}{z} + 3\),
\(Y_1(z) = 1 + \frac{1}{z}\),
\(Y_0(z) = \frac{e^{-z}}{z}\),
and \(z = m_\pi r_{12}\).

Finally, given these expressions for the magnetic dipole moment operator in coordinate space,
we can apply the regulator scheme consistent with the interaction.
Since there are no contact terms in these currents we only need to multiply
these coordinate-space expressions by the regulator \(f(r_{12}/R)\) from \eqref{eq:local-regulator-function}.
In Appendix~\ref{sec:cons-semi-local} we demonstrate the consistency of the SCS-regularized current.

Note that these operators are written involving products of the basic vector
operators $\uvec{r}_{12}$, $\vec{\sigma}_1$, and $\vec{\sigma}_2$.  In order to calculate two-body matrix
elements of these operators, it is advantageous to carry out angular momentum
recoupling on these products to break these operators into spherical tensor
components with definite total orbital angular momentum $L$ and definite total
spin angular momentum $S$. We have provided such tensor decompositions in
Appendix~\ref{sec:tensor-decomposition}.

\section{NCSM calculations for the three-nucleon system}
\label{sec:calculations}

The $A=3$ ground state wave functions, for which we deduce magnetic dipole
moments in the present work, are obtained from \textit{ab initio} no-core shell
model (NCSM)~\cite{barrett-2013-ab, vary-2018-effective} calculations with the
LENPIC interactions.
In the NCSM approach we start with an \(A\)-body Hamiltonian of the form:
\begin{maybemultline}
  \label{eq:ncsm-hamiltonian}
  H
  = \frac{1}{2 m_N A} \sum_{i < j}^A (\vec{p}_i - \vec{p}_j)^2
\ifproofpre{\\}{}
  + \sum_{i < j}^A V_{\mathrm{2N}, ij} + \sum_{i < j < k}^A V_{\mathrm{3N}, ijk}
  + \cdots,
\end{maybemultline}
where the terms on the right hand side are the relative kinetic energy, $2N$
interactions, and $3N$ interactions, respectively. The many-body nuclear wave
functions \(\ket{\Psi}\) are the eigenstates of this Hamiltonian, obtained by
solving the \(A\)-body Schr\"{o}dinger equation:
\begin{equation}
  \label{eq:a-body-scrhodinger}
  H\ket{\Psi} = E\ket{\Psi},
\end{equation}
where \(E\) is the energy eigenvalue corresponding to the state \(\ket{\Psi}\).

The
wave functions are expanded in an complete orthonormal basis
\(\{\ket{\Phi}\}\), where the basis states \(\ket{\Phi}\) are Slater determinants of
single-particle states \(\ket{\phi}\) occupied by the system's nucleons, with
fixed parity and fixed total angular momentum projection.  That is,
\begin{equation}
  \label{eq:many-body-basis}
  \ket{\Phi} = \mathcal{A} \Biggl[\prod_{i=1}^A \ket{\phi_{\alpha_i}}\Biggr],
\end{equation}
where the label \(\alpha_i\) denotes the quantum numbers of nucleon \(i\), and
\(\mathcal{A}\) is the antisymmetrization operator.  The three-dimensional
harmonic oscillator (HO) basis, characterized by the energy parameter
\(\hw\), is the conventional choice for the single particle basis, which we
adopt here.

The resulting many-body basis \(\{\ket{\Phi}\}\) is, in principle, infinite, but, for
actual calculations, we must truncate it.  In the usual \(N_{\mathrm{max}}\)
truncation scheme, configurations are selected by limiting the total number of
HO quanta, shared among the nucleons, to \(N_{\mathrm{max}}\), relative to the
minimum number of quanta required by the Pauli principle.  This truncation
scheme, in particular, ensures a well-behaved center-of-mass wave function
(\textit{e.g.}, Ref.~\cite{caprio2020:intrinsic}).

Expressed in terms of the many-body basis, the \(A\)-body Schr\"{o}dinger
equation~\eqref{eq:a-body-scrhodinger} becomes a finite-dimensional matrix
eigenproblem, where the matrix elements of the Hamiltonian are defined as
\(\braket{\Phi_{\nu}|H|\Phi_{\mu}}\) with \(\mu\) and \(\nu\) labeling the many-body basis states.  The exact result, corresponding to the full, untruncated
many-body problem, is recovered in the limit \(N_{\mathrm{max}} \to
\infty\). Furthermore, given a large enough \(N_{\mathrm{max}}\) the expectation
value of an observable computed in these bases will approach independence of
\(\hw\). We use the Many Fermion Dynamics for nucleons (MFDn)
package~\cite{maris-2010-scaling, shao-2018-accelerating-nuclear} to solve this
matrix eigenvalue problem and obtain the ground state energies and
corresponding many-body wave functions of the $3N$ systems. We then
compute the magnetic dipole moment $\mu(J)$ for these many-body state wave functions,
using the magnetic dipole moment operator ${\vec{\mu}}$ considered above in Sec.~\ref{sec:derivation}.\footnote{In terms of this operator, the
  magnetic dipole moment $\mu(J)$~\cite{suhonen-2007-nucleons-to-nucleus} of a many-body state of angular momentum $J$
  is defined as the expectation value $\mu(J)\equiv\tme{JJ}{\mu_{z}}{JJ}$ of the
  $z$ component in the stretched ($M=J$) substate.   Equivalently, in terms of the reduced matrix
  element~\cite{varshalovich-1988-quantum-theory} of ${\vec{\mu}}$,
$\mu(J)=(2J+1)^{-1/2}\tcg{J}{J}{1}{0}{J}{J}\trme{J}{{\vec{\mu}}}{J}$,
where $\tcg{j_1}{m_1}{j_2}{m_2}{J}{M}$ is a Clebsch-Gordan coefficient.}

For the $A=3$ nuclei, calculations can readily be carried out to sufficiently
high \(\Nmax\), with the LENPIC \(2N\) potentials, to yield the magnetic dipole moment with a numerical precision
which is effectively unlimited.
However, the present
calculations are also intended to explore the use of \(\chi\)EFT currents with
the NCSM in anticipation of future application throughout the range of nuclei
accessable to the NCSM.  In general, the accessable $\Nmax$ may be expected to
critically limit precision which can be obtained for magnetic dipole
observables.

Although the \(\chi\)EFT interaction at \NnLO{2}\ includes $3N$ contributions,
incorporating these $3N$ into NCSM calculations adversely impacts the sparsity
of the many-body Hamiltonian matrix in the NCSM basis, typically imposing an
order-of-magnitude penalty in computational
demands~\cite{maris2013:ncci-chiral-ccp12}.  Thus, the sensitivity of the
calculated magnetic observables to the $3N$ interaction are not only of physical
interest but also of computational interest.  We calculate magnetic dipole
moments for $A=3$ wave functions obtained from the \NnLO{2} LENPIC
interaction, including either only the $2N$ contributions to this interaction
(LENPIC $2N$) or also the $3N$ contributions (LENPIC $2N+3N$).

Furthermore, in calculations for all but the very lightest nuclei, in order to
provide reasonable convergence for accessible values of $\Nmax$, the ``bare''
LENPIC interaction must typically be softened.  This is accomplished by applying
a similarity renormalization group (SRG)
transformation~\cite{glazek-1993-renormalization-hamiltonians,
  glazek-1994-perturbative-renormalization, wegner-1994-flow,
  bogner-2007-similarity-renormalization, bogner-2008-convergence,
  jurgenson-2011-evolving, jurgenson-2013-structure, epelbaum-2019-few-many}. In
the SRG approach, the Hamiltonian in a suitable representation (e.g., here,
momentum representation) is evolved to a band-diagonal structure by a continuous
unitary transformation
\begin{math}
H(\alpha) = U(\alpha)H(\alpha = 0)U^{\dagger}(\alpha),
\end{math}
where \(H(\alpha = 0)\) is the starting Hamiltonian, and \(\alpha\) is the flow
parameter that characterizes the transformation.  Applying this transformation
to a Hamiltonian with $2N$ interactions induces $3N$ and higher many-body
interactions, although the induced interactions are typically truncated at the
$3N$ level.  The impact of SRG transformation on calculated dipole moments in
NCSM calculations, even if such SRG evolution is not actually necessary in the
$A=3$ case, is thus of interest.

Applying a unitary transformation to the Hamiltonian necessitates that the same
transformation be applied to operators for observables.  SRG evolution of a $2N$
current operator may be expected to induce $3N$ (and higher many-body)
contributions to the current operator as well.  Here we restrict ourselves to probing the error
incurred by applying a typical SRG transformation ($\alpha\lesssim \qty{0.1}{\femto\meter\tothe{4}}$) to the
Hamiltonian, without considering the induced corrections to the magnetic dipole
operator.  We carry out calculations in which either the LENPIC $2N$ or LENPIC $2N + 3N$
interactions are SRG evolved, in both cases retaining induced interactions up to
$3N$.

Before then extracting a magnetic dipole moment from the resulting wave function, we must
specify the values for the masses and LECs that appear in the current operator
[see~\eqref{eq:mu-1N-NLO}, \eqref{eq:mu-2N-NLO-Rr-cmdep}, and
  \eqref{eq:mu-2N-NLO-Rr-cmindep}]. For masses and low-energy constants (LECs) which appear in
the expressions both for the potentials~\cite{epelbaum-2015-improved,
  epelbaum-2015-precision-nucleon} and for the magnetic dipole operator, we use the
values already adopted for the potentials: \(m_{N} =
\qty{938.919}{\mega\electronvolt}\), \(m_{\pi} =
\qty{138.03}{\mega\electronvolt}\), \(F_{\pi} = \qty{92.4}{\mega\electronvolt}\)
and \(g_{A} = 1.29\). For the isoscalar and isovector $g$ factors of the
nucleon, we have used \(g_{s} = 0.880\) and \(g_{v} = 4.706\).

\section{Results and discussion}
\label{sec:results}

\begin{figure*}[tp]
  \begin{center}
  \includegraphics[width=\textwidth]{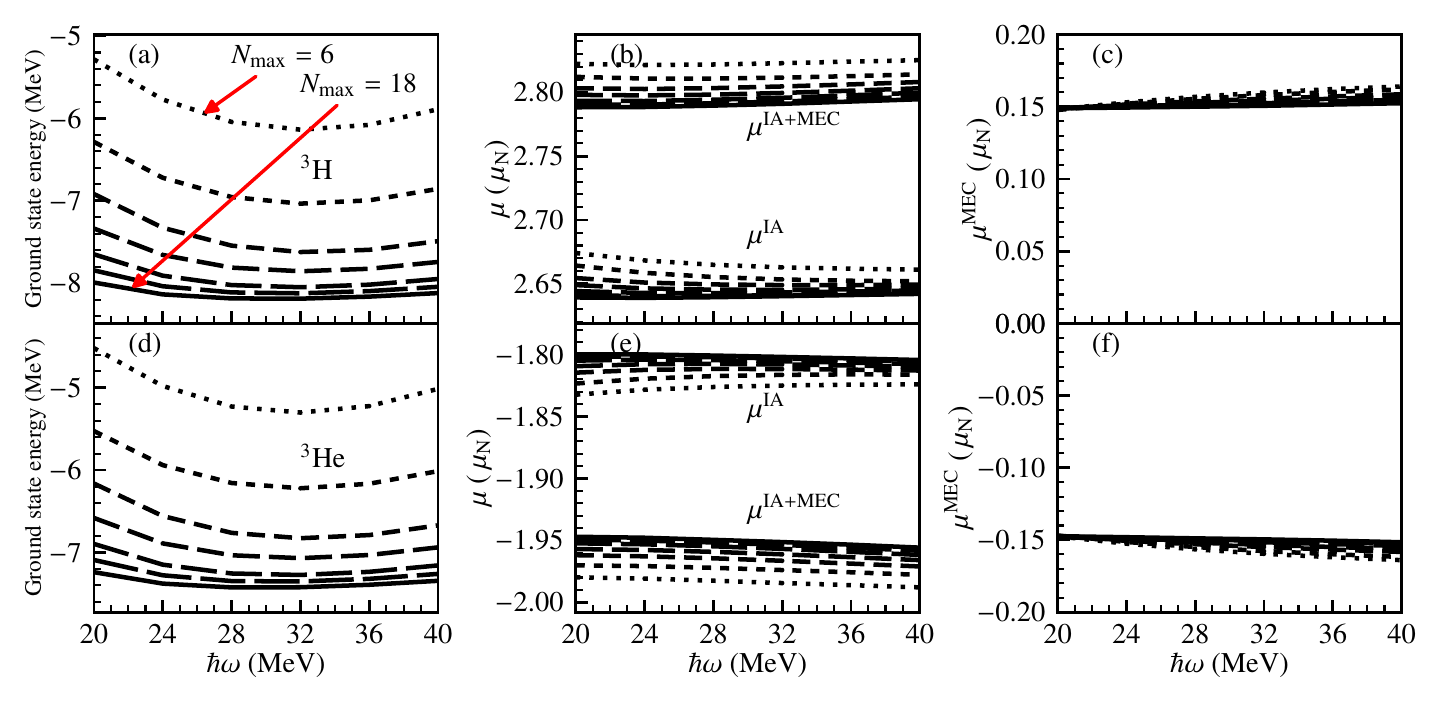}
  \end{center}
  \caption{\label{fig:mu-convergence-trinucleon} Calculated ground state
    energies~(left), magnetic dipole moments~(center), and $2N$ MEC
    corrections (right), for $\isotope[3]{H}$~(top) and
    $\isotope[3]{He}$~(bottom), illustrating their convergence with respect to
    basis parameters \(\Nmax\) (successive curves) and $\hw$. For the magnetic
    dipole moment, \(\mu^{\text{IA}}\) represents the contribution from the
    \(1N\) impulse approximation (IA) operator \(\vec{\mu}^{1N}\) in
    equation~\eqref{eq:mu-1N-NLO}, \(\mu^{\text{MEC}}\) represents the
    contribution from the \(2N\) MEC operator
    \(\vec{\mu}^{2N}\) in equation~\eqref{eq:mu-2N-NLO-Rr}, and \(\mu^{\text{IA}
      + \text{MEC}} = \mu^{\text{IA}} + \mu^{\text{MEC}}\).  Wave functions are
    obtained using the $2N$ LENPIC SCS potential with \(R =
    \qty{1.0}{\femto\meter}\) and no SRG transformation of the potential.}
\end{figure*}

\begin{table*}[tp]
  \centering
  \caption{\label{tab:trinucleon-magnetic-moments}Magnetic moments of the $A=3$
    nuclides calculated with the LENPIC potentials and currents, with consistent
    LECs, at \NnLO{2} (the last nonvanishing contribution
    to the current thus arises at NLO). The SRG-unevolved LENPIC $2N$ calculations
    are shown for $\Nmax=18$, while the other LENPIC calculations are shown for
    $\Nmax=14$, with $\hw$ based on the variational energy minimum.
    Estimated uncertainties from basis truncation
    are discussed in the text.  Prior results obtained with INOY
    ($\Lambda=\qty{500}{\mega\electronvolt}$ and
    $\qty{900}{\mega\electronvolt}$)~\cite{song-2009-up},
    AV18+IL7~\cite{pastore-2013-quantum-monte}, and Norfolk
    (NV2+3-IIb\textsuperscript{*})~\cite{schiavilla-2019-local} potentials, with
    the \(\chi\)EFT current taken to NLO, are shown for comparison, as are the
    experimental values~\cite{nds2015:003}.  The SRG parameter $\alpha$ is given in units of $\unit{\femto\meter\tothe{4}}$, $\hw$ is in units of $\unit{\mega\electronvolt}$, and the magnetic moment in units of \unit{\nuclearmagneton}.
    }
  \begin{ruledtabular}
    \begin{tabular}{lcclccclccc}
      & & & \multicolumn{4}{c}{$\isotope[3]{H}$} & \multicolumn{4}{c}{$\isotope[3]{He}$}\\
      \cmidrule(l){4-7} \cmidrule(l){8-11}
      Potential & \(R\) (\(\unit{\femto\meter}\)) & \(\alpha\) & \(\hw\) & \(\mu^{\text{IA}}\) & \(\mu^{\text{MEC}}\) & \(\mu^{\text{IA} + \text{MEC}}\) &  \(\hw\) & \(\mu^{\text{IA}}\) & \(\mu^{\text{MEC}}\) & \(\mu^{\text{IA} + \text{MEC}}\) \\
      \midrule
      LENPIC $2N$
      & 0.9 & 0.00 & 36 & 2.629 & 0.173 & 2.802 & 36 & -1.796 & -0.172 & -1.968\\
      & 1.0 & 0.00 & 32 & 2.640 & 0.151 & 2.791 & 28 & -1.801 & -0.149 & -1.950\\
      \midrule
      LENPIC $2N$
      & 1.0 & 0.04 & 20 & 2.677 & 0.143 & 2.820 & 16 & -1.822 & -0.141 & -1.963\\
      \mbox{\quad + induced $3N$}
      & 1.0 & 0.08 & 14 & 2.692 & 0.139 & 2.831 & 14 & -1.831 & -0.138 & -1.969\\
      \midrule
      LENPIC $2N$+$3N$
      & 1.0 & 0.04 & 20 & 2.667 & 0.147 & 2.814 & 20 & -1.817 & -0.145 & -1.962\\
      & 1.0 & 0.08 & 14 & 2.683 & 0.142 & 2.825 & 14 & -1.827 & -0.141 & -1.968\\
      \midrule
      \multicolumn{3}{l}{INOY (\NLO; $\qty{500}{\mega\electronvolt}$)\textsuperscript{a}}
      &     & 2.657 & 0.103 & 2.760 &     & -1.810 & -0.103 & -1.913\\
      \multicolumn{3}{l}{INOY (\NLO; $\qty{900}{\mega\electronvolt}$)\textsuperscript{a}}
      &     & 2.657 & 0.172 & 2.829 &     & -1.810 & -0.170 & -1.980\\
      \multicolumn{3}{l}{AV18+IL7 (\NLO)\textsuperscript{a}}
      &     & 2.556 & 0.253 & 2.809 &     & -1.743 & -0.248 & -1.991\\
      \multicolumn{3}{l}{Norfolk (\NLO)\textsuperscript{a}}
      &     & 2.588 & 0.227 & 2.815 &     & -1.770 & -0.224 & -1.994\\
      \midrule
      Experiment
      &     &     &   &    &       & 2.979 &    &    &        & -2.128
    \end{tabular}
  \end{ruledtabular}
  \flushleft \textsuperscript{a} The tabulated values for prior calculations are partial results
  calculated with the \(\chi\)EFT current taken to NLO, and thus involve the same
  diagrams as appear in the current operator used in calculating the present LENPIC results.
\end{table*}

\begin{figure*}[p]
  \begin{center}
  \includegraphics[width=1.0\textwidth]{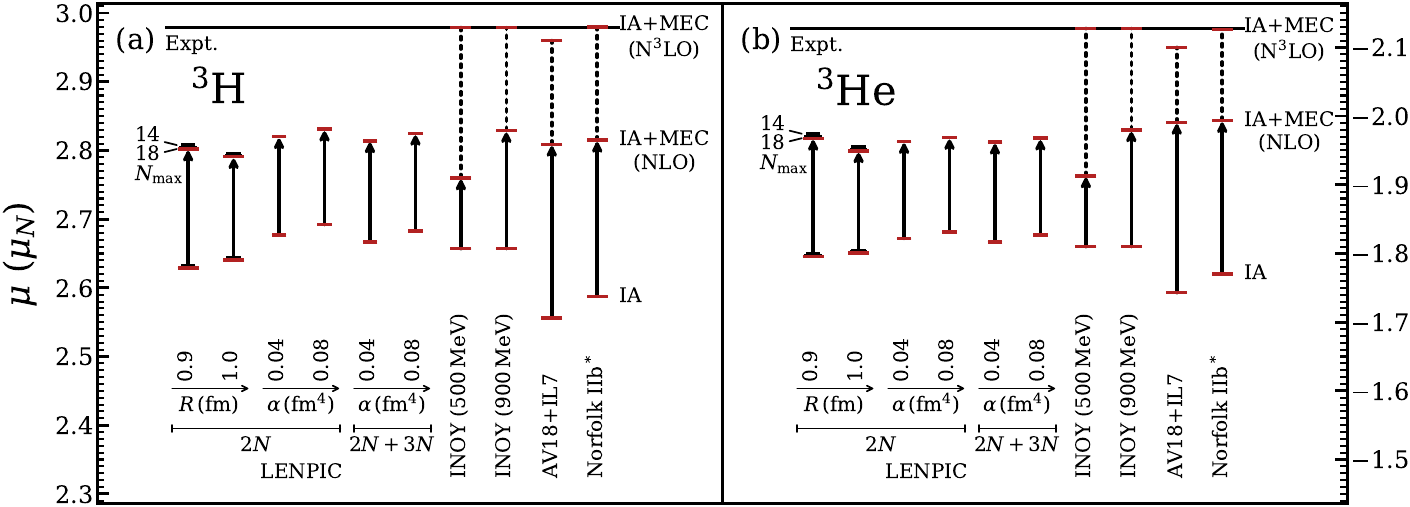}
  \end{center}
  \caption{\label{fig:mu-teardrop} Magnetic moments of the $A=3$ nuclides
    (a)~$\isotope[3]{H}$ and (b)~$\isotope[3]{He}$, calculated with the
    LENPIC potentials and currents, along with results of prior
    calculations~\cite{song-2009-up,pastore-2013-quantum-monte,schiavilla-2019-local}
    (see Table~\ref{tab:trinucleon-magnetic-moments} caption for details of
    these calculations).  Both IA and IA+MEC results are shown (connected by
    arrow), where the \(\chi\)EFT current contains contributions from terms
    appearing through NLO.  For the prior calculations, results including
    contributions through \NnLO{3} are also shown (connected by dotted line);
    however, these results involve new LECs which are chosen to replicate (at
    least approximately) the experimental $A=3$ moments, and are thus not
    predictions \textit{per se} (see text).  The SRG-unevolved LENPIC $2N$
    results are shown for $\Nmax=14$, $16$, and $18$ (increasing symbol size),
    as an indicator of convergence.  The SRG-unevolved LENPIC $2N$ results are
    shown for both \(R = \qty{0.9}{\femto\meter}\) and \(R =
    \qty{1.0}{\femto\meter}\), while all other LENPIC results are shown for \(R
    = \qty{1.0}{\femto\meter}\).  Experimental values~\cite{nds2015:003} are
    provided for reference (horizontal bars).  Note that the magnetic dipole
    moment axis for $\isotope[3]{He}$~(right) is inverted, to facilitate
    comparison of the pattern of MEC contributions (of approximately equal
    magnitude but opposite sign, as noted in the text) across mirror nuclides.
  }
\end{figure*}

Considering first the bare, SRG-unevolved LENPIC $2N$ interaction, convergence
patterns are shown in Fig.~\ref{fig:mu-convergence-trinucleon} for the
calculated ground state energy~(left), magnetic dipole moment~(center), and
$2N$ meson exchange current (MEC) correction (right), for both $\isotope[3]{H}$~(top) 
and $\isotope[3]{He}$~(bottom).  In particular, these calculations are for SCS
regulator parameter \(R = \qty{1.0}{\femto\meter}\).  We carry out these NCSM
calculations, for the $2N$ interaction, through $\Nmax=18$, with \(\hbar\omega\)
from \(\qty{20}{\mega\electronvolt}\) to \(\qty{40}{\mega\electronvolt}\) in
steps of \(\qty{4}{\mega\electronvolt}\).
Note that the variational minimum of the
calculated energies [Fig.~\ref{fig:mu-convergence-trinucleon}~(left)] occurs
within this range.

Calculated dipole moments are shown [Fig.~\ref{fig:mu-convergence-trinucleon}~(center)],
as they are obtained with just the $1N$ impulse-approximation (IA)
dipole operator (\(\mu^{\text{IA}}\)) or including the $2N$ \(\chi\)EFT
corrections as well (\(\mu^{\text{IA} + \text{MEC}}\)).  Both of these
contributions to the moment arise from terms in the current operator which
appear at NLO, while it may be recalled (from Sec.~\ref{sec:derivation}) that
the $1N$ contribution to the current arising at \NnLO{2} does not contribute to
the magnetic dipole moment.  The difference between these curves thus represents
the total MEC correction through \NnLO{2}
[Fig.~\ref{fig:mu-convergence-trinucleon}~(right)].

Numerical results for the calculated dipole moments are tabulated in
Table~\ref{tab:trinucleon-magnetic-moments}, as obtained at the highest
\(\Nmax\) and at the \(\hw\) corresponding to the approximate location of the
variational mininum of the ground state energy on our \(\hw\) mesh starting from $\qty{14}{\mega\electronvolt}$.
These same values for the calculated dipole moments are
summarized graphically in Fig.~\ref{fig:mu-teardrop}, to facilitate comparison
while reading the following discussion.

The approach to numerical convergence in the calculated dipole moment is
evidenced in Fig.~\ref{fig:mu-convergence-trinucleon}~(center), as curves for
successive $\Nmax$ become compressed against each other and as the $\hw$
dependence tends to decrease for the curves of higher $\Nmax$.  Taking $\isotope[3]{H}$
[Fig.~\ref{fig:mu-convergence-trinucleon}~(top)] for illustration, at the
variational energy minimum ($\hw\approx\qty{32}{\mega\electronvolt}$), the IA
moments [Fig.~\ref{fig:mu-convergence-trinucleon}~(b)] for $\Nmax=14$ and
$\Nmax=16$ differ by $\qty{0.002}{\nuclearmagneton}$, and those for $\Nmax=16$
and $\Nmax=18$ differ by only $\qty{0.0009}{\nuclearmagneton}$. The variation of
the IA moment with $\hw$ at $\Nmax=18$, over an interval extending by
$\qty{4}{\mega\electronvolt}$ to each side ($\qty{28}{\mega\electronvolt}\leq
\hw \leq \qty{36}{\mega\electronvolt}$) is $\qty{0.0019}{\nuclearmagneton}$.
The basis dependence of the MEC correction
[Fig.~\ref{fig:mu-convergence-trinucleon}~(c)] is similar on an absolute scale
(e.g., the calculated corrections for $\Nmax=16$ and $\Nmax=18$ differ again by
$\qty{0.0009}{\nuclearmagneton}$, and at $\Nmax=18$ the MEC correction is
nearly independent of $\hw$), and both the IA and MEC contributions thus
contribute similarly to the basis dependence of the calculated total (IA+MEC)
moment [Fig.~\ref{fig:mu-convergence-trinucleon}~(b)].
However, the basis dependence of the IA and the MEC correction is such that the combined result, \(\mu^{\text{IA} + \text{MEC}}\), exhibits a weak but seemingly persistent $\hw$ dependence over the \(\qty{20}{\mega\electronvolt}\) window shown in Fig.~\ref{fig:mu-convergence-trinucleon}~(center), even though it does seem to converge with $\Nmax$.  We can therefore not put a firm numerical uncertainty on our calculated magnetic moments.

The calculated MEC contributions for the mirror nuclides $\isotope[3]{H}$
[Fig.~\ref{fig:mu-convergence-trinucleon}~(c)] and $\isotope[3]{He}$
[Fig.~\ref{fig:mu-convergence-trinucleon}~(f)] are approximately
equal in magnitude ($\qty{0.15}{\nuclearmagneton}$) but opposite in sign.  This
is to be expected as a consequence of isospin symmetry, given that the sole MEC
contribution at NLO may be seen, from the isospin factor
in~(\ref{eq:mu-2N-NLO-Rr}), to be manifestly isovector.  To facilitate comparison
of the pattern of MEC contributions (of approximately equal magnitude but
opposite sign) across the mirror nuclides, note that the magnetic dipole
moment axis in Fig.~\ref{fig:mu-teardrop}~(b) is inverted.  The MEC
contribution provides an $\approx6\%$ correction to the IA moment for
$\isotope[3]{H}$, or $\approx8\%$ for $\isotope[3]{He}$.  In each case, the
correction serves to increase the magnitude of the moment, providing a positive
correction to the positive $\isotope[3]{H}$ moment and negative correction to
the negative $\isotope[3]{He}$ moment.

The experimentally observed dipole moments, for comparison, are
$\qty{2.979}{\nuclearmagneton}$ for $\isotope[3]{H}$ and
$\qty{-2.128}{\nuclearmagneton}$ for $\isotope[3]{He}$~\cite{nds2015:003}. In
each case, the IA calculation underpredicts the magnitude of the moment, and the
MEC contribution thus has the sign needed to resolve the discrepancy, but the
size of the correction is only about half that required to provide agreement
with experiment (see Table~\ref{tab:trinucleon-magnetic-moments} and
Fig.~\ref{fig:mu-teardrop}).

Here we may compare with prior results for the $A=3$ system.  Hybrid
calculations, that is, with wave functions obtained from phenomenological
potentials but moments extracted using \(\chi\)EFT currents, were carried out in
Ref.~\cite{song-2009-up} with the INOY $2N$
potential~\cite{doleschall2003:nonlocal-a3-inoy}, using wave functions obtained
from solving the Faddeev equations, and in
Ref.~\cite{pastore-2013-quantum-monte} with the AV18+IL7 $2N+3N$
potentials~\cite{wiringa1995:nn-av18,pieper2001:3n-il2}, using the Green's
function Monte Carlo (GFMC) many-body method~\cite{carlson2015:qmc-nuclear}.
Then, in Ref.~\cite{schiavilla-2019-local}, fully \(\chi\)EFT calculations with
the Norfolk potential and currents were obtained in calculations in a
hyperspherical harmonic basis.

These works carry the current operators to \NnLO{3}, thus including higher-order
contributions than considered in the present work.  However, they also provide a
detailed breakdown of the contributions to the calculated magnetic moments,
arising from terms appearing at different orders in the \(\chi\)EFT current
operator.  Results obtained by retaining only terms through NLO in the MEC
contribution, summarized in Table~\ref{tab:trinucleon-magnetic-moments} and
Fig.~\ref{fig:mu-teardrop}, include the same diagrams as the present MEC results
and are thus directly comparable.  In the INOY calculations~\cite{song-2009-up},
the IA moments are essentially identical to those found here (to within
$\lesssim\qty{0.02}{\nuclearmagneton}$), and the MEC corrections (at NLO) are
comparable in size to those found here (the INOY results obtained for different
choices of regulator cutoff $\Lambda$ bracket the present results).  In both
the AV18+IL7 and Norfolk calculations, the IA moment is modestly smaller in
magnitude than calculated here (by $\lesssim\qty{0.1}{\nuclearmagneton}$).
However, the NLO correction is correspondingly larger than calculated here,
yielding IA+MEC results at NLO closely similar to those obtained here.

The additional MEC contributions appearing up to \NnLO{3} in the \(\chi\)EFT
currents introduce new LECs, which, in the prior
calculations~\cite{song-2009-up,pastore-2013-quantum-monte,schiavilla-2019-local},
were fit so as to reproduce the experimental moments for the $A=3$
nuclei. Thus, the moments from these calculations
[Fig.~\ref{fig:mu-teardrop} (dotted lines)] do not constitute predictions
\textit{per se} and, indeed, match experiment by construction\footnote{In more recent auxiliary field
  diffusion Monte Carlo (AFDMC) calculations~\cite{martin:afdmc-chiral-m1-PREPRINT}, results with such an approach are, alternatively, contrasted with results for a global
  fit of the LECs to moments for a selection of light nuclei.}
(the small
deviation from experiment in the AV18+IL7 results arises due to differences in
the Hamiltonian, as well as in certain other approximations, between these GFMC
calculations and the few-body calculation actually used in fitting the
LECs~\cite{pastore-2013-quantum-monte}).

Having discussed the qualitative features of the results, let us now examine the
sensitivity of the calculated moments in quantitative detail to choices made,
first, in the \(\chi\)EFT regulator scheme (Sec.~\ref{sec:derivation}) and,
subsequently, in the calculational process (Sec.~\ref{sec:calculations}).
Comparing calculations with regulator cutoff length scale of $R =
\qty{0.9}{\femto\meter}$ to those (just considered) with $R =
\qty{1.0}{\femto\meter}$ induces shifts in the IA moment of
$\lesssim\qty{0.006}{\nuclearmagneton}$, and changes to the MEC contribution of
$\lesssim\qty{0.011}{\nuclearmagneton}$ (see
Table~\ref{tab:trinucleon-magnetic-moments} and Fig.~\ref{fig:mu-teardrop}).
It can easily be understood that the MEC contribution is more sensitive to the regulator scale, because both the wave function and the MEC operator depend on the regulator, whereas the IA operator is independent of the regulator.

Simultaneous SRG evolution of both the interaction and the moment operator, with
all induced many-body contributions retained (only operators through $3N$ are
relevant in the $A=3$ system) would leave the results strictly unchanged in the
full, untruncated space for the problem.  We consider calculations with SRG flow
parameter values $\alpha=\qty{0.4}{\femto\meter\tothe{4}}$ and $\qty{0.8}{\femto\meter\tothe{4}}$.  Induced interactions are retained
through $3N$, but only the unevolved moment operator is used.  That is, only the
calculated wave functions differ in these calculations, without compensating
changes to the operator for the observable. This provides an extreme test of
sensitivity to SRG evolution in the calculational scheme.  Numerical results are
tabulated in Table~\ref{tab:trinucleon-magnetic-moments} for the highest
$\Nmax$ calculated, in this case $\Nmax=14$, again for $\hw$ at the approximate
location of the variational mininum of the ground state energy on our \(\hw\) mesh
for each interaction employed.\footnote{ For the SRG-evolved LENPIC \(2N\) and \(2N + 3N\) interactions,
we carry out the NCSM calculations through \(\Nmax=14\), with \(\hbar\omega = 14, 16, 20, 24, 28\), in \(\unit{\mega\electronvolt}\).}
The resulting changes in the calculated IA moments, as a function of the SRG flow
parameter, are $\lesssim\qty{0.05}{\nuclearmagneton}$, and the changes in the
calculated MEC contribution are $\lesssim\qty{0.012}{\nuclearmagneton}$.
Note that these changes are of the same order as, or even larger than, the basis dependence shown in Fig.~\ref{fig:mu-convergence-trinucleon} for the unrenormalized magnetic moments.

Finally, inclusion of the $3N$ contributions to the interaction may in general
be expected to have signficant effects on the structure and on calculated
observables~\cite{epelbaum-2019-few-many}.  We recalculate the $A=3$ wave functions
using the full LENPIC $2N+3N$ interaction, again with SRG flow parameter values
$\alpha=\qty{0.4}{\femto\meter\tothe{4}}$ and $\qty{0.8}{\femto\meter\tothe{4}}$
(see Table~\ref{tab:trinucleon-magnetic-moments}).
However, we find that including the $3N$ interaction appears to have minimal effect ($\lesssim\qty{0.02}{\nuclearmagneton}$) on the moments obtained for these wave functions; however, one should keep in mind that the moment operator was not SRG-evolved, and we cannot exclude the possibility that the effect of $3N$ interaction is larger when consistent SRG evolved operators are used.

\section{Summary}
\label{sec:summary}

In this work, we calculated the magnetic dipole moments of $\isotope[3]{H}$ and
$\isotope[3]{He}$ with a chirally-improved magnetic dipole operator, within the
context of the NCSM.  Starting from the momentum-space representation of the
LENPIC $\chi$EFT vector current, we derived the SCS-regularized magnetic dipole
operator up to \NnLO{2} in chiral order (the methods presented here generalize
to higher chiral and multipole orders).  We then performed consistent
calculations of magnetic dipole moments of these nuclei, with the semilocal
coordinate-space regularized LENPIC $2N$ and LENPIC $2N + 3N$ potentials. Here, by a
``consistent calculation'' we mean that we adopt both the operators and the
nuclear potentials up to the same chiral order in the calculation.

This work represents our first step towards consistent calculations of
electromagnetic observables using \(\chi\)EFT currents and LENPIC interactions
with the NCSM framework.  Our results are similar to those of prior
theoretical calculations~\cite{song-2009-up, pastore-2013-quantum-monte,
  schiavilla-2019-local}, when taken with the corresponding NLO current operator, likewise falling
short of the experimental values of the magnetic dipole moments of both
$\isotope[3]{H}$ and $\isotope[3]{He}$ by \(6\%\) -- \(8\%\).
Including higher order currents will be essential for a
more comprehensive description of nuclear systems (beyond $A=3$) within the NCSM
framework.

\begin{acknowledgments}
  We would like to thank H.~Krebs, J.~Golak, R.~Skibinski, G.~B.~King, and
  S.~Pastore for useful discussions and sharing numerical results of their
  studies. This material is based upon work supported by the
  U.S.~Department of Energy, Office of Science, under Award
  Nos.~DE-FG02-87ER40371, DE-FG02-95ER40934,  DE-SC0018223
  (SciDAC-4/NUCLEI), DE-SC0023495 (SciDAC-5/NUCLEI), and No. DE-SC0023692. 
  This research used resources of the National Energy Research Scientific Computing
  Center (NERSC), a U.S. Department of Energy Office of Science User Facility located
  at Lawrence Berkeley National Laboratory, operated under Contract 
  No.~DE-AC02-05CH11231 using NERSC award NP-ERCAP0020944.
\end{acknowledgments}

\onecolumngrid
\appendix

\section{Electromagnetic multipole operators \label{sec:em-mult-operators}}

Here we present a general method to derive all electromagnetic multipole
operators from any \(a\)-nucleon (\(\aN{a}\)) charge or current derived from
\(\chi\)EFT. This is a generalization of the procedure described in section
\ref{sec:derivation}. Electric multipole operators are derived from the charge, and
magnetic multipole operators are derived from the current.

\subsection{Electric multipole operators \label{sec:electr-mult-oper}}

To derive electric multipole operators from an \(\aN{a}\) charge
operator we adopt the following definition \cite{bohr-1998-nuclear-structure,
  carlson-1998-structure}:
\begin{align}
  \label{eq:El-definition-1}
  \vec{E}^{\aN{a}}_{l, \order}
  (\vec{r}_1', \ldots, \vec{r}_a', \vec{r}_1, \ldots, \vec{r}_a)
  \equiv \int \dl^3 \vec{x}\, x^{l} \vec{Y}_{l}(\hat{\vec{x}})
  \bar{\rho}^{\aN{a}}_{\order}
  (\vec{r}_1', \ldots, \vec{r}_a', \vec{r}_1, \ldots, \vec{r}_a, \vec{x}),
\end{align}
where \(\bar{\rho}^{\aN{a}}_{\order}\) is the coordinate space representation of the
charge, and \(l\) is the order of the multipole operator.
Note that in this appendix, and the following Appendix~\ref{sec:cons-semi-local}, we
follow the alternative normalization convention of Refs.~\cite{kolling2009:tpe-current,koelling-2011-two} for the expressions for
momentum-space matrix elements, in which~\eqref{eq:moment-space-normalization-a-body} becomes
\begin{equation}
  \label{eq:moment-space-normalization-a-body-old}
    \tme{\vec{p}'_1 \cdots \vec{p}'_a}{{\vec{j}}(\vec{k})}{\vec{p}_1 \cdots \vec{p}_a} 
    = \delta^{(3)}(\vec{q}_1+\cdots+\vec{q}_a - \vec{k})
\vec{j}(\vec{q}_1, \ldots, \vec{q}_a, \vec{Q}_1, \ldots, \vec{Q}_a; \vec{k}),
\end{equation}
and the Fourier transform~\eqref{eq:current-fourier-transformation} relating the expressions for momentum-space and coordinate-space matrix elements becomes
\begin{multline}
  \label{eq:current-fourier-transformation-old}
  \bar{\vec{j}}^{\aN{a}}_{\order}(\vec{r}'_1, \ldots, \vec{r}'_a, \vec{r}_1, \ldots, \vec{r}_a; \vec{x})
    = \intd_{\{\vec{q}\}_1^{a}}
    \intd_{\{\vec{Q}\}_1^{a}}
    \intd_{\vec{k}}
  \prod_{i=1}^a
    e^{i\vec{q}_i \cdot (\vec{r}_i' + \vec{r}_i) / 2}\,
    e^{i\vec{Q}_i \cdot \Delta\vec{r}_i}\,
    e^{i\vec{k} \cdot \vec{x}}\,
  \ifproofpre{\\ \times}{}
    \vec{j}^{\aN{a}}_{\order}(\vec{q}_1, \ldots, \vec{q}_a, \vec{Q}_1, \ldots, \vec{Q}_a; \vec{k})
    \times
    \bar{\delta}^{(3)}(\vec{q}_1 + \cdots + \vec{q}_a - \vec{k}),
\end{multline}
where again \(\intd_{\{\vec{q}\}_{1}^{a}} = \intd_{\vec{q}_{1}} \cdots
\intd_{\vec{q}_{a}}\), but now with \(\intd_{\vec{q}} = \int \frac{\dl^3
  q}{(2\pi)^{3}}\), and \(\bar{\delta}^{(3)}(\cdots) = (2\pi)^3\delta^{(3)}(\cdots)\).
Then the
relation between \(\bar{\rho}^{\aN{a}}_{\order}\), and the momentum space
representation \(\rho^{\aN{a}}_{\order}\) (derived in \cite{krebs2019:n4lo-em,
  koelling-2011-two}) is the same as the relation between
\(\vec{\bar{j}}^{\aN{a}}_{\order}\) and \(\vec{j}^{\aN{a}}_{\order}\) in
equation~\eqref{eq:current-fourier-transformation-old} above.

We simplify equation
(\ref{eq:El-definition-1}) using the following identity
\cite{varshalovich-1988-quantum-theory}:
\begin{align}
  \label{eq:xl-Yl-identity-1}
  x^{l} \vec{Y}_{l}(\unit{x})
  = \sqrt{\frac{(2l + 1)!!}{4\pi l!}} \bigl[\ldots\bigl[\vec{x}
  \vec{x}\bigr]_2 \vec{x}\bigr]_3 \ldots \vec{x}\bigr]_{l}.
\end{align}
With this identity equation (\ref{eq:El-definition-1}) becomes
\begin{multline}
  \label{eq:El-definition-2}
    \vec{E}_{l}^{\aN{a}}
    =(-i)^l
    \sqrt{\frac{(2l + 1)!!}{4\pi l!}}
    \Biggl[
    \bigl[\ldots\bigl[\grad_{\vec{k}} \grad_{\vec{k}}\bigr]_2
    \grad_{\vec{k}}\bigr]_3 \ldots \grad_{\vec{k}}\bigr]_{l}
    \\
    \intd_{\vec{q}_1 \cdots \vec{q}_a} \intd_{\vec{Q}_1 \cdots \vec{Q}_a}
    \prod_{i=1}^a e^{i\vec{q}_i \cdot (\vec{r}_i' + \vec{r}_i) / 2}\,
    e^{i\vec{Q}_i \cdot \Delta\vec{r}_i}\,
    \rho^{\aN{a}}
    \bar{\delta}^{(3)}(\vec{q}_1 + \cdots + \vec{q}_a - \vec{k})
    \Biggr]_{\vec{k}=0},
\end{multline}
where we used
\(\grad_{\vec{k}} e^{i\vec{k}\cdot\vec{x}} =
i\vec{x}e^{i\vec{k}\cdot\vec{x}}\). The tensor product is interpreted as first
applying the \(\grad_{\vec{k}}\)s to the integral and then extracting the
required irreducible tensor component from the result. This is best understood
with an example. The electric quadrupole operator, modulo conventional factors,
is
\begin{align}
  \label{eq:E2-definition}
  & \vec{E}_{2}^{\aN{a}}
    =-\sqrt{\frac{15}{8\pi}}
    \Biggl[
    \bigl[\grad_{\vec{k}} \grad_{\vec{k}}\bigr]_2
    \intd_{\vec{q}_1 \cdots \vec{q}_a} \intd_{\vec{Q}_1 \cdots \vec{Q}_a}
    \prod_{i=1}^a
    e^{i\vec{q}_i \cdot (\vec{r}_i' + \vec{r}_i) / 2}\,
    e^{i\vec{Q}_i \cdot \Delta\vec{r}_i}\,
      \rho^{\aN{a}}
      \bar{\delta}^{(3)}(\vec{q}_1 + \vec{q}_2 - \vec{k})
    \Biggr]_{\vec{k}=0}.
\end{align}
The leading order (LO) $1N$ charge is
\(\rho^{\aN{1}}_{\mathrm{\LO}} (\vec{q}_1, \vec{Q}_1, \vec{k}) = (e / 2)(1 +
\tau_1^3) (2\pi)^3 \delta^{(3)}(\vec{q}_1 - \vec{k})\). Substituting this in the
above equation gives us
\begin{align}
  \label{eq:E2-1N}
  \vec{E}_{2, \mathrm{\LO}}^{\aN{1}}(\vec{r}_1', \vec{r}_1)
  = \Biggl(
  -\sqrt{\frac{15}{8\pi}}
  \bigl[\grad_{\vec{k}} \grad_{\vec{k}}\bigr]_2
  e^{i\vec{k}\cdot\vec{r}_1} \Big|_{\vec{k}=0}
  \Biggr)
  \frac{e}{2}(1 + \tau_1^3) \delta^{(3)}(\vec{r}_1' - \vec{r}_1).
\end{align}
Each of the \(\grad_{\vec{k}}\) acting on \(e^{i\vec{k}\cdot\vec{r}_1}\) will
bring down an \(i\vec{r}_1\). Extracting the rank-2 irreducible tensor from the
resulting tensor we get \(\sqrt{15/8\pi}\bigl[\vec{r}_1 \vec{r}_2\bigr]_2\),
which following equation (\ref{eq:xl-Yl-identity-1}) is simply
\(r_1^2 \vec{Y}_2(\vec{\hat{r}}_1)\). Thus the LO $1N$
electric quadrupole operator is
\begin{align}
  \label{eq:E2-1N-final}
  \vec{E}_{2, \mathrm{\LO}}^{\aN{1}}(\vec{r}_1', \vec{r}_1)
  = \frac{e}{2}(1 + \tau_1^3)
  r_1^2 \vec{Y}_2(\vec{\hat{r}}_1) \delta^{(3)}(\vec{r}_1' - \vec{r}_1),
\end{align}
which is equivalent to the impulse approximation definition of the electric
quadrupole moment operator found in nuclear physics textbooks
\cite{suhonen-2007-nucleons-to-nucleus, bohr-1998-nuclear-structure}.

\subsection{Magnetic multipole operators \label{sec:magn-mult-oper}}

Generalizing equation (\ref{eq:mu-general-def}) we define the \(m\)th spherical
component of the rank \(l\) magnetic multipole operator as
\begin{align}
  \label{eq:Ml-definition-1}
  M_{lm, \order}^{\aN{a}}
    (\vec{r}_1, \ldots, \vec{r}_a, \vec{r}_1', \ldots, \vec{r}_a)
    \equiv
    \frac{1}{l + 1}
    \int \dl^{3} \vec{x} [\vec{x} \times \bar{\vec{j}}^{\aN{a}}_{\order}
    (\vec{r}_1', \ldots, \vec{r}_a', \vec{r}_1, \ldots, \vec{r}_a, \vec{x})]
    \cdot
    \grad[x^l Y_{lm}(\unit{x})],
\end{align}
where \(\bar{\vec{j}}^{\aN{a}}_{\mathcal{O}}\) has been defined in
(\ref{eq:current-fourier-transformation}), and \(Y_{lm}(\unit{r})\) is the
\(m\)th spherical component of \(\vec{Y}_{l}(\unit{r})\). We use the following
identity to simplify this equation \cite{varshalovich-1988-quantum-theory}:
\begin{align}
  \label{eq:xl-Yl-grad-identity-1}
  \grad[x^l Y_{lm}(\unit{x})]
  = \sqrt{l(2l + 1)}
  x^{l-1} \vec{Y}_{lm}^{l-1}(\unit{x}),
\end{align}
where \(\vec{Y}_{lm}^{n}\) is a vector spherical
harmonic whose \(\nu\)th spherical component is given by
\begin{align}
  \label{eq:vector-Yl-component}
  (\vec{Y}_{lm}^{n})_{\nu}
  = (-1)^\nu C^{lm}_{n\, m+\nu\, 1\, \nu} Y_{n-1\,m-\nu}.
\end{align}
With this identity, equation (\ref{eq:xl-Yl-identity-1}) and the definition of
the tensor product equation (\ref{eq:Ml-definition-1}) becomes
\begin{align}
  \label{eq:Ml-definition-2}
  & M_{lm, \order}^{\aN{a}}
    = \frac{\sqrt{l(2l + 1)}}{l + 1}
    \sqrt{\frac{(2l - 1)!!}{4\pi(l - 1)!}}
    \int \dl^{3} \vec{x}
    \bigl[
    \bigl[\cdots\bigl[\vec{x} \vec{x}\bigr]_2
    \vec{x}\bigr]_3 \cdots \vec{x}\bigr]_{l-1}
    \bigl[\vec{x} \times \bar{\vec{j}}^{\aN{a}}_{\order}\bigr]
    \bigr]_{lm},
\end{align}
where for brevity we dropped the arguments of the current. Since this equation
is true for all projections \(m\), we can drop the projection index and write
the above equation as a tensor equation. Again employing
\(\grad_{\vec{k}} e^{i\vec{k}\cdot\vec{x}} = i\vec{x}e^{i\vec{k}\cdot\vec{x}}\)
we get the following final form for the magnetic multipole operators:
\begin{multline}
    \label{eq:Ml-definition-3}
    \vec{M}_{l, \order}^{\aN{a}}
    = (-i)^{l}
    \frac{\sqrt{l(2l + 1)}}{l + 1}
    \sqrt{\frac{(2l - 1)!!}{4\pi(l - 1)!}}
    \Biggl[
    \bigl[
    \bigl[\bigl[\ldots\bigl[\grad_{\vec{k}} \grad_{\vec{k}}\bigr]_2
    \grad_{\vec{k}}\bigr]_3 \ldots \grad_{\vec{k}}\bigr]_{l-1}
    \\
    \bigl[
    \grad_{\vec{k}} \times
    \intd_{\vec{q}_1 \cdots \vec{q}_a} \intd_{\vec{Q}_1 \cdots \vec{Q}_a}
    \prod_{i=1}^a e^{i\vec{q}_i \cdot (\vec{r}_i' + \vec{r}_i) / 2}\,
    e^{i\vec{Q}_i \cdot \Delta\vec{r}_i}\,
    \vec{j}^{\aN{a}}_{\order}
\bigr]
    \bigr]_{l}
    \Biggr]_{\vec{k}=0}.
\end{multline}
The interpretation of the tensor products of the \(\grad_{\vec{k}}\)'s is similar
to that in the case of the electric multipole operators. We first apply the
\(\grad_{\vec{k}}\)'s to the integral and then extract the required irreducible
tensor from the result. For \(l = 1\), after multiplying by the conventional
factor \(\sqrt{4\pi/3}\), the above equation reduces to equation
(\ref{eq:mu-general-def-momentum}). 
\section{Consistentency of semi-local coordinate space regularized current}
\label{sec:cons-semi-local}

The consistency of the current is determined by whether it satisfies the
continuity equation:
\begin{align}
  \label{eq:continuity-momentum-1}
  \vec{k} \cdot \hat{\vec{j}} = \bigl[\hat{H}, \hat{\rho}\bigr],
\end{align}
in momentum space or equivalently
\begin{align}
  \label{eq:continuity-coordinate-1}
  \vec{\nabla}_{\vec{x}} \cdot \hat{\vec{j}} = -i\bigl[\hat{H}, \hat{\rho}\bigr],
\end{align}
in coordinate space. (In this appendix, a hat on a symbol denotes an operator.)
Here \(\hat{j}^\mu = \{\hat{\rho}, \hat{\vec{j}}\}\) is the four-current
operator, \(\hat{H} = \hat{T} + \hat{V}\) is the strong part of the nuclear
Hamiltonian where \(\hat{T}\) denotes the kinetic energy, and \(\hat{V} =
\hat{V}_{\mathrm{\LO}} + \hat{V}_{\NLO} + \cdots\) denotes the potential energy,
and the divergence is with respect to the position of the external
electromagnetic source. As discussed in \cite{riska-1984-electromagnetic,
  pastore-2008-electromagnetic, krebs2019:n4lo-em}, the 1N current satisfies
the continuity equation with the kinetic energy, and the first 2N current at
\NLO{} satisfies the continuity equation with the LO unregularized
potential energy. We do not regularize the kinetic energy. We just have to check
the continuity equation for the 2N current.

The momentum space representation of the four-current operator (where, as in
Appendix~\ref{sec:em-mult-operators}, we follow the normalization
conventions of Refs.~\cite{kolling2009:tpe-current,koelling-2011-two}) is~\cite{koelling-2011-two}:
\begin{align}
  \label{eq:current-mom-representation}
  \braket{\vec{p}_1' \vec{p}_2'|
    \hat{j}^{\aN{2}, \mu}(\vec{k})|
  \vec{p}_1 \vec{p}_2}
  = \delta^{(3)}(\vec{q}_1 + \vec{q}_2 - \vec{k})
  j^{\aN{2}, \mu}(\vec{q}_1, \vec{q}_2, \vec{Q}_1, \vec{Q}_2; \vec{k}).
\end{align}
If \(j^{\aN{a}, \mu}\) does not depend on the \(\vec{Q}_i\) (as is
true for the current under consideration), then the 
coordinate space representation is of the form
\begin{equation}
  \label{eq:current-coord-representation}
        \bar{j}^{\aN{2}, \mu}(\vec{r}'_1, \vec{r}'_2, \vec{r}_1, \vec{r}_2; \vec{x})
        =
        \braket{\vec{r}_1' \vec{r}_2'|
          \hat{j}^{\aN{2}, \mu}(\vec{x})|
          \vec{r}_1 \vec{r}_2}
        = \frac{1}{(2\pi)^3}
        \delta^{(3)}(\vec{r}_1' - \vec{r}_1)
        \delta^{(3)}(\vec{r}_2' - \vec{r}_2)
        \bar{j}^{\aN{2}, \mu}(\vec{r}_1, \vec{r}_2; \vec{x}),
\end{equation}
where the relation between \(j^{\aN{2}, \mu}\) and \(\bar{j}^{\aN{2}, \mu}\) is the same as for the three-currents in
equation~(\ref{eq:current-fourier-transformation-old}).
We will use the momentum representation to check the continuity equation for the
\NLO{} 2N current. The left-hand side of equation
\eqref{eq:continuity-momentum-1} in the momentum representation is
\begin{align}
  \label{eq:lhs-continuity-momentum-1}
  \braket{\vec{p}_1'\vec{p}_2' |
  \vec{k} \cdot \hat{\vec{j}}^{\aN{2}}_{\mathrm{\NLO}} | \vec{p}_1\vec{p}_2}
  = \vec{k} \cdot
  \vec{j}^{\aN{2}}_{\mathrm{\NLO}} (\vec{q}_1, \vec{q}_2)
  \delta^{(3)}(\vec{q}_1 + \vec{q}_2 - \vec{k}),
\end{align}
where \(\vec{j}^{\aN{2}}_{\mathrm{\NLO}}\) has been defined in equation
\eqref{eq:j-2N-NLO}. Doing the dot product, while replacing \(\vec{k}\) by
\(\vec{q}_1 + \vec{q}_2\), we get
\begin{align}
  \label{eq:lhs-continuity-momentum-2}
  \vec{k} \cdot \vec{j}^{\aN{2}}_{\mathrm{\NLO}}
  &=
    i \frac{eg_A^2}{4F_\pi^2} (\vec{\tau}_1 \times \vec{\tau}_2)_z
    \biggl(
    \frac{\vec{\sigma}_1 \cdot \vec{q}_1 \vec{\sigma}_2 \cdot \vec{q}_1}
    {q_1^2 + m_\pi^2}
    -
    \frac{\vec{\sigma}_1 \cdot \vec{q}_2 \vec{\sigma}_2 \cdot \vec{q}_2}
    {q_2^2 + m_\pi^2}
    \biggr).
\end{align}
To evaluate the right hand side of \eqref{eq:continuity-momentum-1} we need the
momentum representation of the unregularized LO potential:
\begin{align}
  \label{eq:lo-potential-momentum-representation}
  \braket{\vec{p}_1' \vec{p}_2' | \hat{V}_{\mathrm{LO}} | \vec{p}_1 \vec{p}_2}
  = V_{\mathrm{LO}}(\tfrac{1}{2}(\vec{p}_1' - \vec{p}_2' - \vec{p}_1 + \vec{p}_2))
  \delta^{(3)}(\vec{p}_1' + \vec{p}_2' - \vec{p}_1 - \vec{p}_2),
\end{align}
where \(V_{\mathrm{LO}}(\vec{q})\) is given by
\begin{align}
  \label{eq:lo-potential}
  V_{\mathrm{LO}}(\vec{q})
  = \vec{\tau}_1 \cdot \vec{\tau}_2 W_{1\pi}(\vec{q})
  = -\frac{g_A^2}{4F_\pi^2} \vec{\tau}_1 \cdot \vec{\tau}_2
  \frac{\vec{\sigma}_1 \cdot \vec{q}\, \vec{\sigma}_2 \cdot \vec{q}}
  {q^2 + m_\pi^2}.
\end{align}
Now we can evaluate what will be the momentum space representation of
\(\hat{V}_{\mathrm{LO}}\hat{\rho}_{\mathrm{LO}}\):
\begin{align}
  \label{eq:V-rho-momentum-representation-1}
  \braket{\vec{p}_1' \vec{p}_2' | \hat{V}_{\mathrm{\LO}}\hat{\rho}_{\mathrm{\LO}}
  | \vec{p}_1 \vec{p}_2}
  &= \int \dl^{3}\vec{p}_1'' \dl^{3}\vec{p}_2''
    \braket{\vec{p}_1' \vec{p}_2' | \hat{V}_{\mathrm{\LO}} | \vec{p}_1''
    \vec{p}_2''}
    \braket{\vec{p}_1'' \vec{p}_2'' | \hat{\rho}_{\mathrm{\LO}}
    | \vec{p}_1 \vec{p}_2}
    \nonumber\\
  &= \int \dl^{3}\vec{p}_1'' \dl^{3}\vec{p}_2''\,
    V_{\mathrm{\LO}}
    (\tfrac{1}{2}(\vec{p}_1' - \vec{p}_2' - \vec{p}_1'' + \vec{p}_2''))
    \delta^{(3)}(\vec{p}_1' + \vec{p}_2' - \vec{p}_1'' - \vec{p}_2'')
    \nonumber\\
  &\qquad
    [\rho_{\mathrm{\LO}, 1}\delta^{(3)}(\vec{p}_1'' + \vec{p}_2'' - \vec{p}_1 -
    \vec{p}_2 - \vec{k})\delta^{(3)}(\vec{p}_2'' - \vec{p}_2) +
    (1 \leftrightarrow 2)].
\end{align}
Completing the integrals over \(\vec{p}_1''\), and \(\vec{p}_2''\), we get
\begin{align}
  \label{eq:V-rho-momentum-representation-2}
  \braket{\vec{p}_1' \vec{p}_2' | \hat{V}_{\mathrm{\LO}}\hat{\rho}_{\mathrm{\LO}}
  | \vec{p}_1 \vec{p}_2}
  = [V_{\mathrm{\LO}}(\vec{q}_1)\rho_{\mathrm{\LO}, 1} + (1 \leftrightarrow 2)]
  \delta^{(3)}(\vec{q}_1 + \vec{q}_2 - \vec{k}),
\end{align}
where \(\rho_{\mathrm{\LO}, i} = e(1 + \tau_{i}^3)/2\). We can similarly evaluate
\(\hat{\rho}_{\mathrm{\LO}}\hat{V}_{\mathrm{\LO}}\). Using the commutation
relation
\(\bigl[\vec{\tau}_1 \cdot \vec{\tau}_2, \tau_{1}^3\bigr] = 2i (\vec{\tau}_1
\times \vec{\tau}_2)^3\) we see that
\begin{align}
  \label{eq:potential-charge-commutation}
  &\braket{\vec{p}_1' \vec{p}_2' |
    \bigl[ \hat{V}_{\mathrm{\LO}}, \hat{\rho}_{\mathrm{\LO}} \bigr] |
    \vec{p}_1 \vec{p}_2}
    \nonumber\\
  &\quad
    = \biggl(\bigl[ V_{\mathrm{\LO}}(\vec{q}_1), \rho_{\mathrm{\LO}, 1} \bigr]
    + (1 \leftrightarrow 2)\biggr) \delta^{(3)}(\vec{q}_1 + \vec{q}_2 - \vec{k})
    \nonumber\\
  &\quad
    = i \frac{eg_A^2}{4F_\pi^2} (\vec{\tau}_1 \times \vec{\tau}_2)^3
    \biggl(
    \frac{\vec{\sigma}_1 \cdot \vec{q}_1 \vec{\sigma}_2 \cdot \vec{q}_1}
    {q_1^2 + m_\pi^2}
    -
    \frac{\vec{\sigma}_1 \cdot \vec{q}_2 \vec{\sigma}_2 \cdot \vec{q}_2}
    {q_2^2 + m_\pi^2}
    \biggr) \delta^{(3)}(\vec{q}_1 + \vec{q}_2 - \vec{k}).
\end{align}
Thus
\(\braket{\vec{p}_1' \vec{p}_2' | \vec{k} \cdot
  \hat{\vec{j}}_{\mathrm{\NLO}}^{\aN{2}} | \vec{p}_1 \vec{p}_2} =
\braket{\vec{p}_1' \vec{p}_2' | \bigl[ \hat{V}_{\mathrm{\LO}},
  \hat{\rho}_{\mathrm{\LO}} \bigr] | \vec{p}_1 \vec{p}_2}\), i.e. the \NLO{} 2N
current satisfies the continuity equation with the unregularized LO
potential \cite{riska-1984-electromagnetic, pastore-2008-electromagnetic,
  krebs2019:n4lo-em}. Fourier transforming both sides of this equation we get
the continuity equation satisfied by the current in coordinate space,
\( \braket{\vec{r}_1' \vec{r}_2' | \vec{\nabla}_{\vec{x}} \cdot
  \hat{\vec{j}}^{\aN{2}}_{\mathrm{\NLO}} | \vec{r}_1 \vec{r}_2} =
-i\braket{\vec{r}_1' \vec{r}_2' | \bigl[\hat{V}_{\mathrm{\LO}},
  \hat{\rho}_{\mathrm{\LO}}\bigr] | \vec{r}_1 \vec{r}_2}, \) where
\begin{align}
  \braket{\vec{r}_1' \vec{r}_2' |
  \vec{\nabla}_{\vec{x}} \cdot \hat{\vec{j}}^{\aN{2}}_{\mathrm{\NLO}} |
  \vec{r}_1 \vec{r}_2}
  &= \vec{\nabla}_{\vec{x}} \cdot
    \bar{\vec{j}}^{\aN{2}}_{\mathrm{\NLO}}(\vec{r}_1, \vec{r}_2, \vec{x})
    \delta^{(3)}(\vec{r}_1' - \vec{r}_1) \delta^{(3)}(\vec{r}_2' - \vec{r}_2)
  \\
  \label{eq:lo-potential-coordinate}
  \braket{\vec{r}_1' \vec{r}_2' |
  \hat{V}_{\mathrm{\LO}}(\vec{r}_1, \vec{r}_2) |
  \vec{r}_1 \vec{r}_2}
  &= \vec{\tau}_1 \cdot \vec{\tau}_2 \bar{W}_{1\pi}(\vec{r}_1 - \vec{r}_2)
    \delta^{(3)}(\vec{r}_1' - \vec{r}_1) \delta^{(3)}(\vec{r}_2' - \vec{r}_2),
  \\
  \label{eq:lo-charge-coordinate}
  \braket{\vec{r}_1' \vec{r}_2' |
  \hat{\rho}_{\mathrm{\LO}} |
  \vec{r}_1 \vec{r}_2}
  &= e \biggl(\frac{1 + \tau_{1z}}{2} \delta^{(3)}(\vec{r}_1 - \vec{x})
    + (1 \leftrightarrow 2) \biggr)
    \delta^{(3)}(\vec{r}_1' - \vec{r}_1) \delta^{(3)}(\vec{r}_2' - \vec{r}_2),
\end{align}
where \(\bar{j}^{\aN{2}}_{\mathrm{\NLO}}\) and \(\bar{W}_{1\pi}\) are the Fourier transforms of
\(\vec{j}^{\aN{2}}_{\mathrm{\NLO}}\), and \(W_{1\pi}\), respectively.  With this
form for the coordinate space representation we see that
\begin{align}
  \label{eq:nlo-continuity-coordinate}
  \vec{\nabla}_{\vec{x}} \cdot
  \bar{\vec{j}}^{\aN{2}}_{\mathrm{\NLO}}(\vec{r}_1, \vec{r}_2, \vec{x})
  = e (\vec{\tau}_1 \times \vec{\tau}_2)_z
  \bar{W}_{1\pi}(\vec{r}_1 - \vec{r}_2)
  \bigl[\delta^{(3)}(\vec{r}_1 - \vec{x}) -
  \delta^{(3)}(\vec{r}_2 - \vec{x})\bigr].
\end{align}
Introducing a discrete, complete basis, such as the HO basis, we can write this
continuity equation as a matrix equation
\(\vec{\mathcal{Z}} = \vec{\mathcal{W}}\), where the matrix elements are:
\begin{align}
  \label{eq:continuity-equation-unregularized-matrix}
  \vec{\mathcal{Z}}_{\alpha\beta}
  &= \int \dl^{3}\vec{r}_1 \dl^{3}\vec{r}_2\,
    \phi_{\alpha}(\vec{r}_1, \vec{r}_2)
    \vec{\nabla}_{\vec{x}} \cdot
    \bar{\vec{j}}^{\aN{2}}_{\mathrm{\NLO}}(\vec{r}_1, \vec{r}_2, \vec{x})
    \phi_{\beta}(\vec{r}_1, \vec{r}_2), \\
  \vec{\mathcal{W}}_{\alpha\beta}
  &= e (\vec{\tau}_1 \times \vec{\tau}_2)_z
    \int \dl^{3}\vec{r}_1 \dl^{3}\vec{r}_2\,
    \bigl[\delta^{(3)}(\vec{r}_1 - \vec{x}) -
    \delta^{(3)}(\vec{r}_2 - \vec{x})\bigr]
    \phi_{\alpha}(\vec{r}_1, \vec{r}_2)
    \bar{W}_{1\pi}(\vec{r}_1 - \vec{r}_2)
    \phi_{\beta}(\vec{r}_1, \vec{r}_2),
\end{align}
where \(\{\ket{\alpha} = \ket{\alpha_1, \alpha_2}\}\) is the basis, and
\(\braket{\vec{r}_1 \vec{r}_2 | \alpha} = \phi_{\alpha}(\vec{r}_1, \vec{r}_2)\).
We now introduce two new operators \(\hat{f}\), and \(\hat{g}\) such that
\(\braket{\vec{r}_1'\vec{r}_2' | \hat{f} | \vec{r}_1\vec{r}_2} =
f(\vec{r}_1-\vec{r}_2) \delta^{(3)}(\vec{r}_1'-\vec{r}_1)
\delta^{(3)}(\vec{r}_2'-\vec{r}_2)\), and
\(\braket{\vec{r}_1'\vec{r}_2' | \hat{g} | \vec{r}_1\vec{r}_2} =
g(\vec{r}_1,\vec{r}_2) \delta^{(3)}(\vec{r}_1'-\vec{r}_1)
\delta^{(3)}(\vec{r}_2'-\vec{r}_2)\). Here \(f\), and \(g\) are functions of
nucleon coordinates, with no isospin structure. It can be easily checked that
\(\hat{\vec{j}}^{\aN{2}}_{\mathrm{\NLO}}\) commutes with \(\hat{g}\), and
\(\hat{V}_{\mathrm{\LO}}\) commutes with \(\hat{f}\). We want to see how
\(\hat{f}\) and \(\hat{g}\) are related if we demand that
\(\nabla_{\vec{x}} \cdot \bigl(\hat{\vec{j}}^{\aN{2}}_{\mathrm{\NLO}}\hat{g}\bigr) =
-i\bigl[\hat{V}_{\mathrm{\LO}}\hat{f}, \hat{\rho}_{\mathrm{\LO}}\bigr]\). We will
use the coordinate space relations that we have derived above. In the coordinate
space representation, the left-hand side of this equation evaluates to
\begin{align}
  \label{eq:lhs-regularized-continuity-1}
  \braket{\vec{r}_1'\vec{r}_2' |
  \nabla_{\vec{x}} \cdot \hat{\vec{j}}^{\aN{2}}_{\mathrm{\NLO}} \hat{g} |
  \vec{r}_1\vec{r}_2}
  &= \int \dl^{3}\vec{x}_1\dl^{3}\vec{x}_2
    \braket{\vec{r}_1'\vec{r}_2' |
    \nabla_{\vec{x}} \cdot \hat{\vec{j}}^{\aN{2}}_{\mathrm{\NLO}}  |
    \vec{x}_1\vec{x}_2}
    \braket{\vec{x}_1'\vec{x}_2' | \hat{g} | \vec{r}_1\vec{r}_2}
    \nonumber\\
  &= \nabla_{\vec{x}} \cdot \vec{j}^{\aN{2}}_{\mathrm{\NLO}}(\vec{r}_1, \vec{r}_2,
    \vec{x}) g(\vec{r}_1, \vec{r}_2) \delta^{(3)}(\vec{r}_1'-\vec{r}_1)
    \delta^{(3)}(\vec{r}_2'-\vec{r}_2).
\end{align}
To evaluate the right-hand side of this equation, in the coordinate space
representation, we  first need to evaluate \(\braket{\vec{r}_1'\vec{r}_2' |
  \hat{V}_{\mathrm{\LO}}\hat{f}\hat{\rho}_{\mathrm{\LO}} | \vec{r}_1\vec{r}_2}\).
\begin{align}
  \label{eq:rhs-regularized-continuity-1}
  &\braket{\vec{r}_1'\vec{r}_2' |
    \hat{V}_{\mathrm{\LO}}\hat{f}\hat{\rho}_{\mathrm{\LO}} | \vec{r}_1\vec{r}_2}
    \nonumber\\
  &\quad
    = \int \dl^{3}\vec{x}_1' \dl^{3}\vec{x}_2' \dl^{3}\vec{x}_1 \dl^{3}\vec{x}_2
    \braket{\vec{r}_1'\vec{r}_2' | \hat{V}_{\mathrm{\LO}} | \vec{x}_1'\vec{x}_2'}
    \braket{\vec{x}_1'\vec{x}_2' | \hat{f} | \vec{x}_1\vec{x}_2}
    \braket{\vec{x}_1\vec{x}_2 | \hat{\rho}_{\mathrm{\LO}} | \vec{r}_1\vec{r}_2}
    \nonumber\\
  &\quad
    = V_{\mathrm{\LO}}(\vec{r}_1,\vec{r}_2)
    \rho_{\mathrm{\LO}}(\vec{r}_1,\vec{r}_2,\vec{x})
    f(\vec{r}_1-\vec{r}_2) \delta^{(3)}(\vec{r}_1'-\vec{r}_1)
    \delta^{(3)}(\vec{r}_2'-\vec{r}_2),
\end{align}
where \(V_{\mathrm{\LO}}(\vec{r}_1,\vec{r}_2)\), and
\(\rho_{\mathrm{\LO}}(\vec{r}_1,\vec{r}_2,\vec{x})\), are the expressions in the
right-hand side of equations \eqref{eq:lo-potential-coordinate}, and
\eqref{eq:lo-charge-coordinate}, respectively, modulo the delta functions
involving \(\vec{r}_i'\). We can similarly evaluate
\(\braket{\vec{r}_1'\vec{r}_2' |
\hat{\rho}_{\mathrm{\LO}}\hat{V}_{\mathrm{\LO}}\hat{f} | \vec{r}_1\vec{r}_2}\):
\begin{multline}
  \label{eq:rhs-regularized-continuity-2}
  \braket{\vec{r}_1'\vec{r}_2' |
  \bigl[\hat{V}_{\mathrm{\LO}}\hat{f},\hat{\rho}_{\mathrm{\LO}}\bigr] |
  \vec{r}_1\vec{r}_2}
  = e (\vec{\tau}_1 \times \vec{\tau}_2)_z
    \bar{W}_{1\pi}(\vec{r}_1 - \vec{r}_2)
    \bigl[\delta^{(3)}(\vec{r}_1 - \vec{x}) -
    \delta^{(3)}(\vec{r}_2 - \vec{x})\bigr]
    f(\vec{r}_1, \vec{r}_2)
    \delta^{(3)}(\vec{r}_1'-\vec{r}_1) \delta^{(3)}(\vec{r}_2'-\vec{r}_2).
\end{multline}
Thus according to our demand,
\begin{align}
  \label{eq:regularized-continuity-coordinate}
  &\nabla_{\vec{x}} \cdot \vec{j}^{\aN{2}}_{\mathrm{\NLO}}(\vec{r}_1, \vec{r}_2,
    \vec{x}) g(\vec{r}_1, \vec{r}_2) \delta^{(3)}(\vec{r}_1'-\vec{r}_1)
    \delta^{(3)}(\vec{r}_2'-\vec{r}_2)
    \nonumber\\
  &\quad
    = e (\vec{\tau}_1 \times \vec{\tau}_2)_z
    \bar{W}_{1\pi}(\vec{r}_1 - \vec{r}_2)
    \bigl[\delta^{(3)}(\vec{r}_1 - \vec{x}) -
    \delta^{(3)}(\vec{r}_2 - \vec{x})\bigr]
    f(\vec{r}_1, \vec{r}_2)
    \delta^{(3)}(\vec{r}_1'-\vec{r}_1) \delta^{(3)}(\vec{r}_2'-\vec{r}_2).
\end{align}
Then by equation \eqref{eq:nlo-continuity-coordinate} we have
\(\nabla_{\vec{x}} \cdot \vec{j}^{\aN{2}}_{\mathrm{\NLO}}(\vec{r}_1, \vec{r}_2,
\vec{x}) (g(\vec{r}_1, \vec{r}_2) - f(\vec{r}_1 - \vec{r}_2)) = 0\), which means
if \(\nabla_{\vec{x}} \cdot \vec{j}^{\aN{2}}_{\mathrm{\NLO}}\) is not zero, then
\(g(\vec{r}_1, \vec{r}_2) = f(\vec{r}_1 - \vec{r}_2)\). In the discrete,
complete basis, introduced earlier this translates to
\begin{align}
  \label{eq:continuity-equation-regularized-basis}
  \vec{\mathcal{Z}}\vec{\mathcal{G}} = \vec{\mathcal{W}}\vec{\mathcal{F}},
\end{align}
where
\begin{align}
  \label{eq:F-G-matrix}
  \vec{\mathcal{F}}_{\alpha\beta}
  &= \int \dl^3\vec{r}_1\dl^3\vec{r}_2
    \phi_\alpha(\vec{r}_1, \vec{r}_2) f(\vec{r}_1 - \vec{r}_2)
    \phi_\beta(\vec{r}_1, \vec{r}_2),
    \nonumber\\
  \vec{\mathcal{G}}_{\alpha\beta}
  &= \int \dl^3\vec{r}_1\dl^3\vec{r}_2
    \phi_\alpha(\vec{r}_1, \vec{r}_2) g(\vec{r}_1, \vec{r}_2)
    \phi_\beta(\vec{r}_1, \vec{r}_2),
\end{align}
are the matrices corresponding to \(\hat{f}\) and \(\hat{g}\), respectively, in
the discrete, complete basis. Since \(\vec{\mathcal{Z}} = \vec{\mathcal{W}}\),
if \(\vec{\mathcal{Z}}\) is non-singular then
\(\vec{\mathcal{G}} = \vec{\mathcal{F}}\), i.e. the matrix elements of the two
regulators must be the same. Even though here we derived this result for the
2N current at \NLO{}, it is evident that as long as equation
(\ref{eq:nlo-continuity-coordinate}) is satisfied for a particular pair of
current and unregularized potential, we will reach the same conclusion.
 
\section{Tensor decomposition of two-nucleon operators \label{sec:tensor-decomposition}}
We define the rank \(j\) tensor product of two irreducible tensors
\(\vec{T}_{j_1}\) and \(\vec{T}_{j_2}\) of ranks \(j_1\) and \(j_2\),
respectively, as
\begin{align}
  \label{eq:tensor-product-def}
  \bigl[\vec{T}_{j_1} \vec{T}_{j_2}\bigr]_{j}^{m}
  = \sum_{m_1, m_2} C_{j_1m_1j_2m_2}^{jm}
  T_{j_1}^{m_1} T_{j_2}^{m_2},
\end{align}
where \(m\), \(m_1\), and \(m_2\) are the projection indices and
\(C_{j_1m_1j_2m_2}^{jm}\) is a Clebsch-Gordan coefficient. The tensor product
itself is an irreducible tensor. It follows from the above definition and
properties of the Clebsch-Gordan coefficients that the following recoupling
identities hold for commuting tensors \cite{varshalovich-1988-quantum-theory}:
\begin{align}
  \label{eq:commuting-tensor-product-identity-1}
  \bigl[\vec{T}_{a} \vec{T}_{b}\bigr]_{c}
  &= (-1)^{a + b - c}
    \bigl[\vec{T}_{b} \vec{T}_{a}\bigr]_{c},
\end{align}
\begin{align}
  \label{eq:commuting-tensor-product-identity-2}
  & \bigl[\bigl[\vec{T}_{a} \vec{T}_{b}\bigr]_{c}\vec{T}_{d}\bigr]_{e}
    = (-1)^{a + b + d + e}
    \sum_{f} \Pi_{c f}
    \begin{Bmatrix}
      a & b & c \\
      d & e & f
    \end{Bmatrix}
              \bigl[\vec{T}_{a} \bigl[\vec{T}_{b} \vec{T}_{d}\bigr]_{f}\bigr]_{e},
\end{align}
\begin{align}
  \label{eq:commuting-tensor-product-identity-3}
  & \bigl[\bigl[\vec{T}_{a} \vec{T}_{b}\bigr]_{c}
  \bigl[\vec{T}_{d} \vec{T}_{e}\bigr]_{f}\bigr]_{i}
    = \sum_{hi} \Pi_{cfgh}
    \begin{Bmatrix}
      a & b & c \\
      d & e & f \\
      g & h & i
    \end{Bmatrix}
  \bigl[\bigl[\vec{T}_{a} \vec{T}_{d}\bigr]_{g}
  \bigl[\vec{T}_{b} \vec{T}_{e}\bigr]_{h}\bigr]_{i},
\end{align}
where
\(\Pi_{ab \cdots \lambda} = \sqrt{(2a + 1)(2b + 1) \cdots (2\lambda + 1)}\), and
the quantities in the braces in equations
(\ref{eq:commuting-tensor-product-identity-2}) and
(\ref{eq:commuting-tensor-product-identity-3}) are the Wigner \(6j\) and
\(9j\) symbols, respectively. These identities hold for any projections of the
tensor product, hence we have omitted any explicit projection index. Using
\(\vec{T} \cdot \vec{S} = -\sqrt{3}\bigl[\vec{T}_1\vec{S}_1\bigr]_{0}\) and
\(\vec{T} \times \vec{S} = -i\sqrt{2}\bigl[\vec{T}_1\vec{S}_1\bigr]_{1}\), and
the above identities we have
\begin{align}
  \label{eq:spin-tensor-decomposition-1}
  \bigl(\vec{\sigma}_1 \times \vec{\sigma}_2\bigr)
  \cdot \vec{\hat{r}}_{12} \vec{\hat{r}}_{12}
  &= i\frac{\sqrt{2}}{3}
    \Bigl\{
    \vec{\Sigma}_1
    + \sqrt{10} \bigl[
    \vec{C}_{2, 0}^{2}
    \vec{\Sigma}_1\bigr]_1\Bigr\},
\end{align}
\begin{align}
  \label{eq:spin-tensor-decomposition-2}
  \vec{\hat{R}}_{12} \times \vec{\hat{r}}_{12}
  \vec{\sigma}_1 \cdot \vec{\sigma}_2
  &= -i\sqrt{6}
    \bigl[
    \vec{C}_{1, 1}^{1}
    \vec{\Sigma}_0\bigr]_1,
\end{align}
\begin{multline}
  \label{eq:spin-tensor-decomposition-3}
  \vec{\hat{R}}_{12} \times \vec{\hat{r}}_{12}
    (\vec{\sigma}_1 \cdot \vec{\hat{r}}_{12}
    \vec{\sigma}_2 \cdot \vec{\hat{r}}_{12})
    = i\frac{\sqrt{2}}{3}
    \Biggl\{
    -\sqrt{3}\bigl[\vec{C}_{1, 1}^1 \vec{\Sigma}_0\bigr]_1
    + \sqrt{\frac{3}{5}}\bigl[\vec{C}_{1, 1}^1 \vec{\Sigma}_1\bigr]_1
    + \sqrt{\frac{9}{5}}\bigl[\vec{C}_{1, 1}^2 \vec{\Sigma}_2\bigr]_1
    \\
    + \sqrt{\frac{14}{5}}\bigl[\vec{C}_{3, 1}^2 \vec{\Sigma}_2\bigr]_1
    + \sqrt{\frac{24}{5}}\bigl[\vec{C}_{3, 1}^3 \vec{\Sigma}_2\bigr]_1
    \Biggr\},
\end{multline}
where \(\vec{\Sigma}_l = \bigl[\vec{\sigma}_1 \vec{\sigma}_2\bigr]_l\)
and
\(\vec{C}_{a, b}^c = \bigl[\vec{C}_{a}(\unit{r}_{12})
\vec{C}_b(\unit{R}_{12})\bigr]_{c}\). The
\(\vec{C}_{l}(\unit{r}) = \sqrt{4\pi/(2l + 1)} \vec{Y}_{l}(\unit{r})\) is the
rank \(l\) renormalized spherical harmonic. We would like to remind the reader
that the numerical subscripts associated with the Pauli matrices and the unit
vectors represent nucleon indices and not tensor ranks. Combining all of this we
can write equation (\ref{eq:mu-2N-NLO-Rr}) as
\begin{align}
  \label{eq:mu-2N-NLO-Rr-tensor-decomp}
  \vec{\mu}^{\aN{2}}_{\mathrm{\NLO}}
  &= \frac{2}{3} g_{\pi} \bigl[\vec{\tau}_1 \vec{\tau}_2\bigr]_1
    \Bigl[
    \vec{\mu}'^{\,\aN{2}}_{\mathrm{\NLO, cm-dep}}(\vec{R}_{12}, \vec{r}_{12})
    +
    \vec{\mu}'^{\,\aN{2}}_{\mathrm{\NLO, cm-indep}}(\vec{r}_{12})
    \Bigr]
    \delta^{(3)}(\Delta\vec{R}_{12})
    \delta^{(3)}(\Delta\vec{r}_{12}),
\end{align}
where
\begin{multline}
  \label{eq:mu-2N-NLO-Rr-cmdep-tensor-decomp}
    \vec{\mu}'^{\,\aN{2}}_{\mathrm{\NLO, cm-dep}}(\vec{R}_{12}, \vec{r}_{12})
    =
    (m_{\pi} R_{12})
    \Biggl[
    -\sqrt{3} z \bigl[\vec{C}_{1, 1}^{1} \vec{\Sigma}_0\bigr]_1
    + Y_2(z)
    \Biggl(
    + \sqrt{\frac{3}{5}}\bigl[\vec{C}_{1, 1}^1 \vec{\Sigma}_1\bigr]_1
    + \sqrt{\frac{9}{5}}\bigl[\vec{C}_{1, 1}^2 \vec{\Sigma}_2\bigr]_1
    \\
    + \sqrt{\frac{14}{5}}\bigl[\vec{C}_{3, 1}^2 \vec{\Sigma}_2\bigr]_1
    + \sqrt{\frac{24}{5}}\bigl[\vec{C}_{3, 1}^3 \vec{\Sigma}_2\bigr]_1
    \Biggr)
    \Biggr]Y_0(z),
\end{multline}
and
\begin{align}
  \label{eq:mu-2N-NLO-Rr-cmindep-tensor-decomp}
  \vec{\mu}'^{\,\aN{2}}_{\mathrm{\NLO, cm-indep}}(\vec{r}_{12})
  &= \sqrt{10} (1 + z) \bigl[\vec{C}_{2, 0}^{2} \vec{\Sigma}_{1}\bigr]_1
    + (-1 + 2z) \vec{\Sigma}_{1}.
\end{align} \twocolumngrid

\bibliographystyle{apsrev4-2}

\end{document}